\documentclass[journal]{IEEEtran}
\usepackage{times}
\usepackage{stfloats}
\usepackage{graphicx}
\usepackage{amsmath}
\usepackage{amssymb}
\usepackage{amsthm}
\usepackage{mathtools,algorithm}
\usepackage{bm}
\usepackage{xcolor}
\usepackage{multirow}
\usepackage[T1]{fontenc}
\usepackage{algpseudocode}% 
\usepackage[pagebackref=true,breaklinks=true,colorlinks,bookmarks=false,citecolor=blue]{hyperref}

\newcommand{\ltl}[8]{$\texttt{R}_{\texttt{#1}}\texttt{C}_{\texttt{#2}}\texttt{M}_{\texttt{#3}}\texttt{S}_{\texttt{#4..#5}}\texttt{B}_{\texttt{#6..#7}}\texttt{N}_{\texttt{#8}}$}

\begin{document}
\title{CAT: Cellular Automata on Tensor cores}
\author{Crist\'obal A. Navarro, Felipe A. Quezada, Enzo Meneses, H\'ector Ferrada, Nancy Hitschfeld
\thanks{Crist\'obal A. Navarro, Felipe A. Quezada, Enzo Meneses, and H\'ector Ferrada are with the Department of Informatics of Austral University of Chile and the Temporal research group
(\url{http://temporal.uach.cl)}.}% <-this % stops a space
\thanks{Nancy Hitschfeld is with the Computer Science Department of University of Chile.}% <-this % stops a space
}

\maketitle

\begin{abstract}
Cellular automata (CA) are simulation models that can produce complex emergent behaviors from simple local rules. Although state-of-the-art GPU solutions are already fast due to their data-parallel nature, their performance can rapidly degrade in CA with a large neighborhood radius. With the inclusion of tensor cores across the entire GPU ecosystem,  interest has grown in finding ways to leverage these fast units outside the field of artificial intelligence, which was their original purpose. 
In this work, we present CAT, a GPU tensor core approach that can accelerate CA in which the cell transition function acts on a weighted summation of its neighborhood. 
CAT is evaluated theoretically, using an extended PRAM cost model, as well as empirically using the Larger Than Life (LTL) family of CA as case studies. The results confirm that the cost model is accurate, showing that CAT exhibits constant time throughout the entire radius range $1 \le r \le 16$, and its theoretical speedups agree with the empirical results. At low radius $r=1,2$, CAT is competitive and is only surpassed by the fastest state-of-the-art GPU solution. Starting from $r=3$, CAT progressively outperforms all other approaches, reaching speedups of up to $101\times$ over a GPU baseline and up to $\sim 14\times$ over the fastest state-of-the-art GPU approach. In terms of energy efficiency, CAT is competitive in the range $1 \le r \le 4$ and from $r \ge 5$ it is the most energy efficient approach. As for performance scaling across GPU architectures, CAT shows a promising trend that if continues for future generations, it would increase its performance at a higher rate than classical GPU solutions. The results obtained in this work put CAT as an attractive GPU approach for scientists that need to study emerging phenomena on CA with large neighborhood radius. 
\end{abstract}

\begin{IEEEkeywords}
Cellular Automata, Tensor Cores, Game of Life, Larger than life, GPU Computing, Energy Efficiency.
\end{IEEEkeywords}

\IEEEpeerreviewmaketitle
\section{Introduction}
Cellular automata (CA) are discrete dynamical systems capable of producing complex emergent phenomena \cite{codd2014cellular}. Originally conceived as abstract mathematical systems, CA have found utility in several fields of science and technology; from artificial life systems \cite{adamatzky2010game,langton1986studying,evans2001larger,rokicki2018life}, sand/fluid simulation \cite{dennunzio2009sand,kier1994cellular,d2012cellular} to highly complex ecological and urban systems \cite{hogeweg1988cellular,batty1997urban}, among many others. In terms of computation, a key property of CA is that they are highly data-parallel, making them ideal candidates for parallel computing.

\begin{figure}[ht!]
    \centering
    \includegraphics[width=0.8\columnwidth]{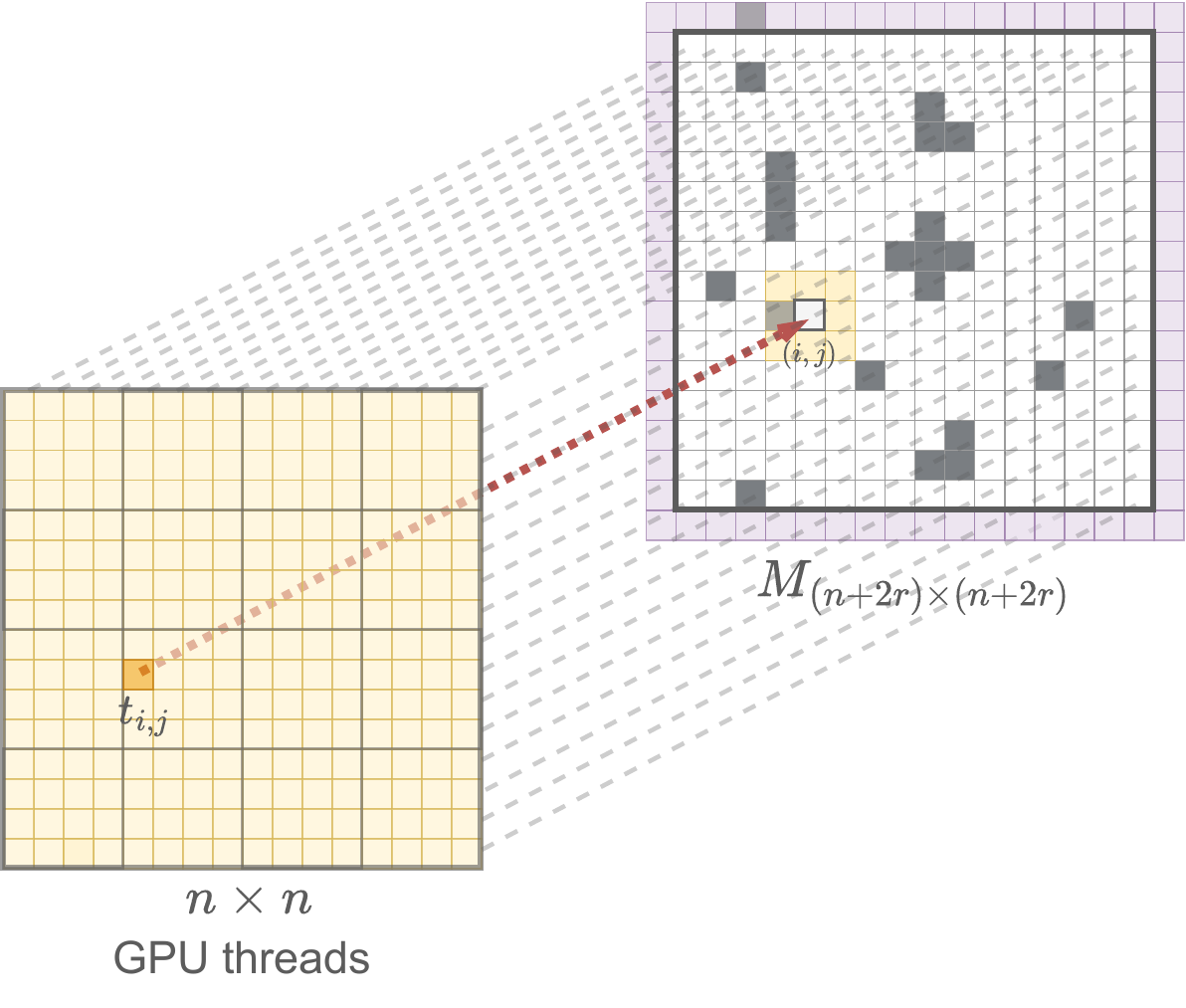}
    \caption{Traditional data-parallel approach for simulating Cellular Automata (CA) using a global halo of ghost cells which is common for avoiding complex logic on the boundary threads. In this example each thread $t_{i,j}$ is in charge of one cell and must explore its Moore neighborhood of radius $r=1$. In general, with this approach one simulation step costs at least $\Omega(r^2)$ time.}
    \label{fig:traditional}
\end{figure}

With the advent of Graphics Processing Units (GPUs), current implementations of CA have harnessed their parallel processing capabilities  \cite{rybacki2009experiments,gibson2015investigation,rumpf2010conways}, enabling the study of larger-scale and more intricate models. State-of-the-art implementations employ several techniques that range from using the programmable cache memory (known as \textit{shared memory} in CUDA) \cite{holewinski2012high}, multi-GPU \cite{zhang2021multi}, to more specific ones such as using multiple cells per thread \cite{millan2017performance}, packet coding \cite{cagigas2022efficient} and multiple step simulation \cite{fujita2015efficient} in programmable cache. These state-of-the-art techniques build on top of the traditional data-parallel scheme for cellular automata where each GPU thread is in charge of one (or more) cells and its entire neighborhood of radius $r$, typically Moore or Von Neumann. An example illustration is shown in Figure \ref{fig:traditional}.

One of the main limitations of the traditional data-parallel scheme is that performance degrades considerably as the CA neighborhood radius $r$ increases. This is because for each thread, a neighborhood of at least $(1+2r) \times (1+2r)$ cells must be explored, producing a memory bound scenario that costs at least $\Omega(r^2)$ memory accesses. This cost per thread limits the possibilities of researching and applying CA models such as Larger than Life \cite{griffeath1994self,evans1996larger,evans2001larger,bekaroglu2023analyzing} which exhibits emergent complex phenomena at large neighborhood radius. 

\begin{figure}[ht!]
    \centering
    \includegraphics[width=0.8\columnwidth]{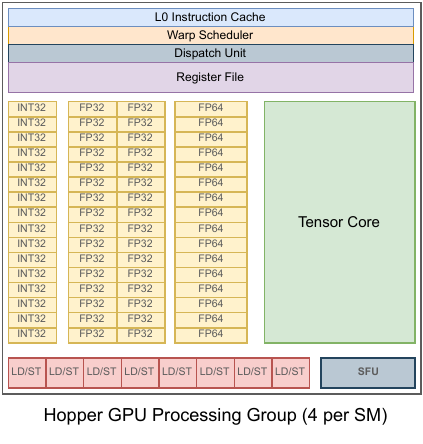}
    \caption{Processing group in Nvidia's Hopper architecture. Four of these make up a streaming multiprocessor (SM), and dozens of SMs form an entire GPU chip. Currently, the number of tensor cores in a GPU chip can reach up to the hundreds. Image inspired from the CUDA C Programming guide \cite{cudaCProgGuide2024}.}
    \label{fig:processing-group}
\end{figure}

In the last decade, GPUs have shifted from pure general purpose parallel processors (\texttt{FP32} / \texttt{FP64} / \texttt{INT32} units), to hybrid ones that also include specific purpose units which are significantly faster for the task they were designed for. One type of specialized units are the tensor cores\footnote{The other specialized unit is the Ray Tracing (RT) core \cite{sanzharov2019examination}} \cite{markidis2018nvidia,choquette2021nvidia,choquette2023nvidia}, which are part of the GPU chip (as shown in Figure \ref{fig:processing-group}) and offer a hardware accelerated Matrix Multiply Accumulate (MMA) function that runs in one GPU cycle. 
These tensor core units were designed to keep up with the demands of artificial intelligence (AI) applications that require training very large models, such as large language models (LLMs) \cite{narayanan2021efficient} or computer vision models (CVM) \cite{wang2022lightseq2}, among many others. Considering the high performance of tensor cores, and the fact that modern GPU chips can contain hundreds of them, it is relevant to explore whether or not tensor cores can be used to accelerate CA simulations beyond current GPU solutions, and mitigate the performance degradation identified when increasing the neighborhood radius.

In this work, we present CAT (\textbf{C}ellular \textbf{A}utomata on \textbf{T}ensor cores); a GPU method that uses tensor cores to simulate Cellular Automata (CA) at different neighborhood radii. The main feature of CAT is that its cost per simulation step remains constant, \textit{i.e.}, $\mathcal{O}(1)$, as long as the radius is within the dimension of the square tensor core matrix. For current GPUs the matrix dimensions are $16 \times 16$, thus the supported range is $1 \le r \le 16$. With this design, CAT avoids the performance degradation found in other state-of-the-art approaches when increasing the radius. In terms of requirements, CAT can accelerate any CA where the cell transition function acts on a weighted summation of the cell's neighborhood.
%$\langle i,j \rangle$, \textit{i.e.}, \cristobal{Revisar simbologia, lambda es CA} a weighted sum of the form $C_{ij}^{t+1} = \mathcal{F}(C_{ij}) = \sum_{k \in \langle i,j \rangle} a_k C_k$. 
%The implementation provided in this work supports a radius of up to $r <= 16$; this range captures the phenomena of several CA applications that employ radius greater than one. Still, if more radius is required, we provide the main guides to extend the idea for even larger radius. 

The rest of the manuscript is organized as follows: Section \ref{sec:related_work} covers related work, including the state of the art approaches for which CAT will be compared with. Section \ref{sec:tc-model} introduces the tensor core programming model. Section \ref{sec:overview_cat} contains CAT's formulation and analysis of theoretical performance. Section \ref{sec:experimental_results} presents experimental results using the Larger than Life (LTL) family of CA at different neighborhood radius. Section \ref{sec:conclusions} discusses the main results and concludes the work.

\section{Related Work}
\label{sec:related_work}
Up to date, several works have provided performance comparisons between state-of-the-art techniques to simulate CA on multicore CPU and GPU \cite{rybacki2009experiments,gibson2015investigation,biggers2011cuda}. In general, it is well known that GPU implementations can run faster than multi-core CPU ones (assuming high-end hardware on both cases); however, it is worth noting that certain CA features may affect the performance of one or the other significantly. For example, it has been found that more arithmetic operations in the transition function favors more GPU parallelism \cite{gibson2015investigation}, while complex neighborhood accesses or avoiding the quiescent states of a CA is a big challenge for GPUs and less for CPUs \cite{oxman2014computational,rybacki2009experiments}. 

As for research on GPU approaches, Fujita \textit{et al.} presented a GPU implementation for accelerating the simulation of game of life (GOL) \cite{gardner1970fantastic} using a multi-step scheme. This scheme, also known as temporal blocking \cite{zhang2023revisiting}, consists of loading tiles of the CA into CUDA's programmable shared memory (one region per CUDA block), and simulating several time steps per kernel call in order to reuse cache memory. During each kernel call, a halo of outdated cells progressively grows from the perimeter of the CA region to the interior, for $t$-steps. The cells that are further inside (not touched by this halo) become valid simulated cells after all the steps. The authors report large speedups of up to two orders of magnitude over a sequential CPU implementation. However, this approach is best suited for radius $r=1$, as the \textit{outdated cells} boundary propagates at a rate of $r$ cells per time step, limiting the effectiveness of the technique at larger radius. The authors did not link any source code in their manuscript.   

Millan \textit{et al.} \cite{millan2017performance} implemented and compared the performance of classic and state-of-the-art GPU approaches for simulating CA, using the game of life (GOL) as case study as well as a compute-bound variant with additional synthetic computation per time step (in order to manifest maximum parallelism in the GPU). Their results showed that with modern GPU architectures (2017+) the classic global memory implementation is often the fastest approach, followed by a \textit{multi-cell} approach \cite{balasalle2012optimizing} which consists of simulating two (or more) cells per thread in order to re-utilize part of the thread's neighborhood exploration effort on the adjacent cell. The authors also consider radius $r > 1$, and show performance results in the range $1 \le r \le 5$. Their results reveal that indeed the running times increase with the radius because the work per cell increases at least as $\Omega(r^2)$. These results support the fact that running efficient CA simulations with larger neighborhood radius is a known challenge that still requires more research \cite{tran2004new}. In the same work the authors also compared their results with a shared memory approach \cite{topa2013using,topa2014cellular} where for each block, the first four rows of threads are in charge of filling the halo of the shared memory tile. They found that a \textit{shared-memory} approach on stencil-like patterns \cite{schafer2011high} such as CA does not necessarily improve the performance anymore as it did in the past. This behavior has also been reported in another work \cite{RENC2022101538} and agrees with our experimental results when comparing global vs shared memory baselines at low radius. This phenomenon can be attributed in part to the L1/L2 caches now being much more effective than before as well as the increased size of the L2 which caches the global memory accesses of threads. Source codes of both the multi-cell and shared memory implementations were made available by Millan \textit{et al.} \cite{millan2017performance} \footnote{The classic global memory approach is also included in the experiments as part of our own baseline implementations.} and were included in our experimental comparison, named MCELL and SHARED respectively. These implementations were extended by our team to support $r \in [1..16]$ and performance optimizations were made for the multi-cell (MCELL) approach. 

Cagigas \textit{et al.} \cite{cagigas2022efficient} proposed an efficient approach for simulating CA using regular GPU Computing, \textit{i.e.}, no use of tensor cores. The core idea is the application of \textit{packet coding}, a technique where each GPU thread reads a 64-bit word from GPU global memory and simulates 8-bit cells codified inside with bitwise operations. Two benefits arise from this approach; i) just one global coalesced memory access is performed for multiple cells, and ii) the computational effort for accessing the neighborhood is shared among the codified cells. The authors report that doing packet coding is between $3\times$ to $4\times$ faster than the baseline GPU approach, and faster than other state of the art techniques such as \textit{lookup-table} (\textit{i.e.}, to precompute possible outcomes of the transition function, and access the table with the cell's current state and neighbor information) and \textit{temporal blocking} \cite{zhang2023revisiting} implemented with the AN5D framework \cite{matsumura2020an5d}. The work was focused on radius $r=1$, however the approach can be extended to larger radius. The authors made their source code available but only working for radius $r=1$, therefore our team extended the implementation, no named PACK, to support radiuses in the range $r \in [1..16]$. 

Regarding the use of tensor math, Zhuang \textit{et al.} proposed a Python deep learning framework based on high-level tensor computation on GPU to accelerate land simulation CA \cite{zhuang2022tensor}. The authors report that by using their approach, they achieve up to $\sim 50\times$ of speedup over a sequential counterpart running on a CPU. No details are given on the use of tensor-cores, neither tests on other well known cellular automata such as game of life (GOL), as the scope of the work is more oriented at formulating the problem in a high-level tensor framework for an specific application.

Regarding the use of GPU tensor cores, Liu \textit{et al.} \cite{liu2022toward} used them to accelerate the finite difference method (FDM) for partial differential equations (PDEs), which is a stencil-like pattern. The authors report average speedups of up to $\sim4.55\times$ over a highly optimized GPU baseline. This work is relevant to mention because it puts a precedent on the benefits of using tensor cores in stencil-like patterns, and it can even be used to simulate cellular automata as well, although it's design is not suited for large neighborhood radius. The main difference this approach has in comparison to the proposed method (CAT) is that its design is based on doing one pair of MMAs per tensor core cell matrix (known as fragment), which is beneficial, but at the cost of having to overlap fragments according to $r$, and apply an \textit{in-halo} scheme. In the \textit{in-halo} scheme, the effective computation inside a fragment is not $100\%$ because the tensor computation of one MMA between the cell and band matrices cannot capture the entire neighborhood at the boundary. Therefore, the cells at the last $r$ levels of the fragment must be discarded. At low radius such as $r=1$ this is not a big issue, as the fragment is usually $16 \times 16$ in size and would get its effective area reduced to $14 \times 14$. But as the radius increases, the \textit{in-halo} would grow to the point of making most of the fragment surface invalid, leaving just a small effective area at the center. This effect penalizes GPU performance significantly as the process would need to compensate with many overlapped fragments in order to cover all cells of the domain. Because of this, such design is best suited for low neighborhood radius such as $r=1,2$. In contrast, CAT (our proposed method) uses the opposite design; an \textit{out-halo} concept where increasing the radius means extending to the outside using three pairs of MMAs per cell fragment. This design change has favorable implications as it allows the fragment MMA operation to be entirely effective even up to radius $r=16$ (and beyond if the source code is extended; more on this in the discussion of Section \ref{sec:conclusions}). Lastly, Chen \textit{et al.} \cite{10.1145/3627535.3638476} recently proposed an efficient way of doing high-precision (FP64) stencil computations on GPU using tensor cores as well. Their matrix layout is different however, as they convert tiles of the input data to matrix rows and convolution tiles to matrix columns, among other technical improvements such as a lookup table to reduce integer operations and \textit{dirty bits} to alleviate bank conflicts in shared memory.

\section{Tensor Core Programming Model}
\label{sec:tc-model}
%CAT (\textbf{C}ellular \textbf{A}utomata on \textbf{T}ensor cores) relies on the Matrix Multiply Accumulate (MMA) operation, provided by the GPU tensor cores, to capture the neighborhood of groups of cells. Each tensor core offers a fast one-cycle product of small matrices and the addition of the result to an accumulator matrix. Consequently, the programming of all tensor cores allows the parallel simulation of a large CA. 
We provide a brief summary of the tensor core programming model in order to better understand the formulation of CAT in Section \ref{sec:overview_cat}.
The following explanations use Nvidia CUDA terminology for tensor cores, nevertheless the core concepts exposed are all shared by GPU manufacturers such as Nvidia, AMD and Intel. 

The main operation of a tensor core is the Matrix Multiply Accumulate (MMA), defined as 
\begin{equation}
    \label{eq:tensor-core-mma}
    D_{p\times q} = A_{p\times k} \times B_{k\times q} + C_{p\times q}
\end{equation}
with $A_{p\times k}, B_{k\times q}$ being the product matrices, $C_{p\times q}$ the accumulator and $D_{p\times q}$ the resulting matrix. If needed, one can set $D=C$ and keep the result in $C$. At the CUDA programming level, these tensor core matrices are named \texttt{fragments}, and their $p\times q\times k$ sizes will depend on the chosen numerical precision (\texttt{FP16}, \texttt{TF32}, \texttt{INT8}, \texttt{INT4}, etc), and in all cases they are small. For example, at \texttt{half} (FP16) precision, there are 256 elements per fragment, set as $p \times q \times k = 16 \times 16 \times 16$. Non-square shapes are available as well, such as $32 \times 8 \times 16$ or $8 \times 32 \times 16$, and even longer shapes at lower precision such as INT4 or BIT. In this work we are interested in the square fragments with the $p\times q \times k = 16 \times 16 \times 16$ setting. These MMA operations defined at the CUDA programming level, when executed they get further subdivided into actual tensor core instructions which are the ones that cost one GPU cycle. For example, in the Ampere architecture, the one-cycle tensor core instruction executes an MMA of size $p\times q \times k = 8\times 4 \times 8$. Therefore, all MMAs defined at the CUDA programming level are automatically subdivided into the needed number of one-cycle tensor core instructions. 

When writing a CUDA Kernel, Nvidia provides access to the tensor core instructions through the C++ namespace \texttt{wmma}. One of the main differences between the tensor core instructions and the regular CUDA instructions is that the former are not per thread, but instead per warp\footnote{In CUDA threads execute in groups of 32, which are named warps.}. That is, when a thread meets a MMA instruction in a CUDA kernel, it automatically synchronizes with the others threads of the same warp to internally cooperate for the execution of the MMA. The same logic applies for the other tensor-core related instructions such as declaring fragments, loading global memory data into a fragment and writing fragment data into global memory; they are all per-warp. 
%In CUDA, these instructions are offered by default as the set of \texttt{WMMA} instructions. 
A full description of CUDA tensor core programming, including the list of fragment types and example codes, is available at the official NVIDIA CUDA C Programming guide \cite{cudaCProgGuide2024}. It is worth clarifying that regular CUDA code and tensor core code can co-exist in the same kernel.   

Designing an efficient tensor-core based solution involves coming up with a GPU kernel where many warps put the input data from global memory into fragments, and launch MMA operations in parallel. All traditional kernel code is executed by the regular GPU cores, while the MMA operations are executed by the tensor cores.

\section{Formulation of CAT}
\label{sec:overview_cat}
One of the most time consuming tasks in simulating many CA is accessing the neighboring cells, which must be done for each cell of the domain and at each time step of the simulation. Such CA are known to be memory-bound because the dominant cost is in the memory accesses (\textit{i.e.}, neighborhood access) rather than in the arithmetic operations of the transition function. A very well known example of memory-bound CA is John Conway's Game of Life \cite{gardner1970fantastic,adamatzky2010game} and its generalization to any radius known as Larger than Life \cite{griffeath1994self,evans1996larger,evans2001larger} which is even more memory-bound due to the larger neighborhood radius. Apart from being memory-bound, these CA also characterize for having a transition function that acts on the weighted summation of each cell's neighborhood. Simulating memory-bound CA in GPU has its challenges because it saturates the memory bandwidth much earlier than the computational capacity of the chip. Traditional GPU solutions often mitigate this problem by doing an efficient use of the memory hierarchy, including the programmable L1 cache known as \textit{shared memory} in CUDA's terminology. The proposed method, \textit{Cellular Automata on Tensor cores} (CAT), handles these types of CA from a different perspective; by adapting the exploration of neighborhood cells as a series of MMA operations executed by the tensor cores, which take one GPU cycle per MMA.

The formulation of CAT relies on a fundamental linear algebra fact which is that the weighted neighborhood summation of a CA cells (except for the boundary ones) can be computed simultaneously with two matrix products between the entire CA domain and a constant band matrix of ones. The result of these two matrix operations returns a matrix where each element is the weighted sum of all of its neighborhood data including itself twice\footnote{The center cell value can later be subtracted by each thread if needed.}. For neighborhood radius $r=1$, the band matrix is tridiagonal, and for general radius the diagonal band has a width of $2r+1$. A frequent approach for handling the neighborhood of boundary cells more efficiently is to include a global halo of ghost cells of width $r$. Although this global halo has an extra memory cost of $\approx 4nr = \mathcal{O}(n)$, it is not significant compared to the problem size $\mathcal{O}(n^2)$. 

\begin{figure}[ht!]
    \centering
    \includegraphics[width=0.97\columnwidth]{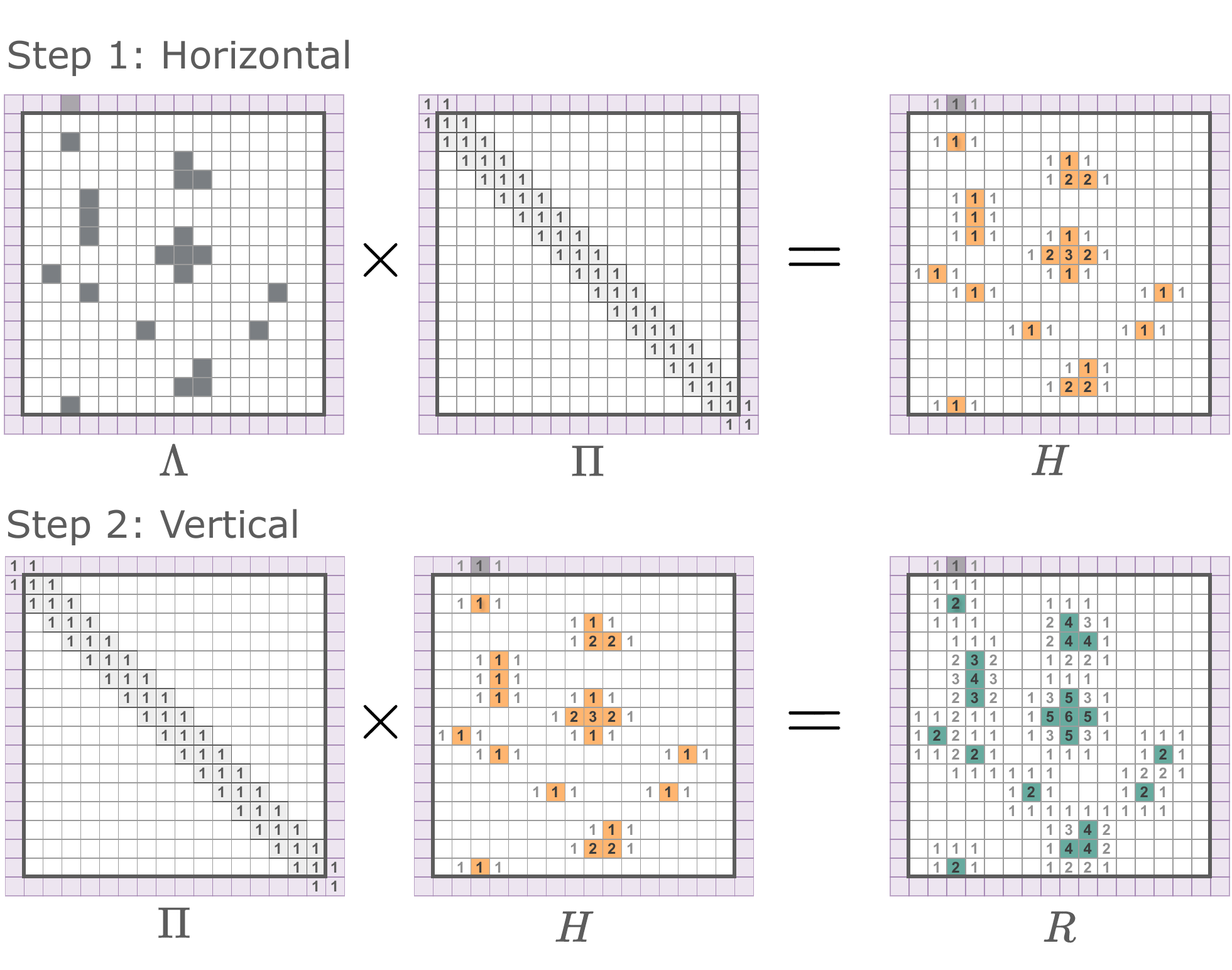}
    \caption{Concept of how a pair of matrix products between an entire CA ($\Lambda$) and a band matrix ($\Pi$) can count the living neighbors of all cells (no tensor core logic introduced yet). Here, the CA includes a global halo of ghost cells, giving a total size of $(n + 2r)\times (n+2r) = 18 \times 18$ as the neighborhood radius is $r=1$. The final cells of $R$ contain their number of living neighbors plus its own state added twice (cells with no number have value zero).}
    \label{fig:cat-theory-example}
\end{figure}

Let $\Lambda_{n' \times n'}$ be the matrix representation of an entire cellular automata of size $n \times n$, using a global halo of ghost cells where $n' = n+2r$, and $\Pi_{n' \times n'}$ the entire band matrix with halo as well, then the reduction of all Moore neighborhoods can be computed with two MMAs:
\begin{align}
    \label{eq:cat-theory-h}
    H_{n' \times n'} &= \Lambda_{n' \times n'} \times \Pi_{n'\times n'} + 0_{n' \times n'}\\
    \label{eq:cat-theory-r}
    R^{\text{Moore}}_{n'\times n'} &= \Pi_{n' \times n'} \times H_{n'\times n'} + 0_{n' \times n'}
\end{align}
The first step reduces horizontally into $H$, while the second step does a vertical reduction using $H$ as input, to end up with the Moore neighborhood reduction matrix $R$. The simplified Von Neumann neighborhood can also be computed, it only requires to redefine Eq. (\ref{eq:cat-theory-r}) as:
\begin{align}
    \label{eq:cat-theory-von-neumann}
    R^{\text{Von Neumann}}_{n'\times n'} &= \Pi_{n' \times n'} \times \Lambda_{n'\times n'} + H_{n' \times n'}
\end{align}
Figure \ref{fig:cat-theory-example} illustrates an example of how the two MMAs (the $0_{n' \times n'}$ matrices were omitted) apply for a Game of Life (GoL) CA of size $n\times n = 16\times 16$ with Moore neighborhood of $r=1$. The purple bands surrounding the $16 \times 16$ domain is the global halo of ghost cells.

In any of the two neighborhood types (Moore or Von Neumann), there will be duplicate state values added on each non-zero cell. If the CA rule needs to exclude the center cell in the neighborhood counting, then one can compute $R - 2\Lambda$ in parallel as an extra instruction in the same GPU kernel, using a standard per-thread logic.  

In order to compute Eqs. (\ref{eq:cat-theory-h}) and (\ref{eq:cat-theory-r}) (or Eq. (\ref{eq:cat-theory-von-neumann}) if needed) efficiently with tensor cores, one has to consider that the products defined between the CA and the band matrices (as shown in Figure \ref{fig:cat-theory-example}), in practice have to be programmed as several tensor core MMAs that occur at a smaller scale and in parallel between fragments of $\Lambda$ and $\Pi$, similar to the blocked matrix multiply scheme. Furthermore, given that the band matrix $\Pi$ is non-zero only on its diagonal band, then the MMAs that actually matter are the ones where the band is present, \textit{i.e.}, all other MMAs can be skipped as the product would compute to zero. 

\begin{figure}[ht!]
    \centering
    \includegraphics[width=0.9\columnwidth]{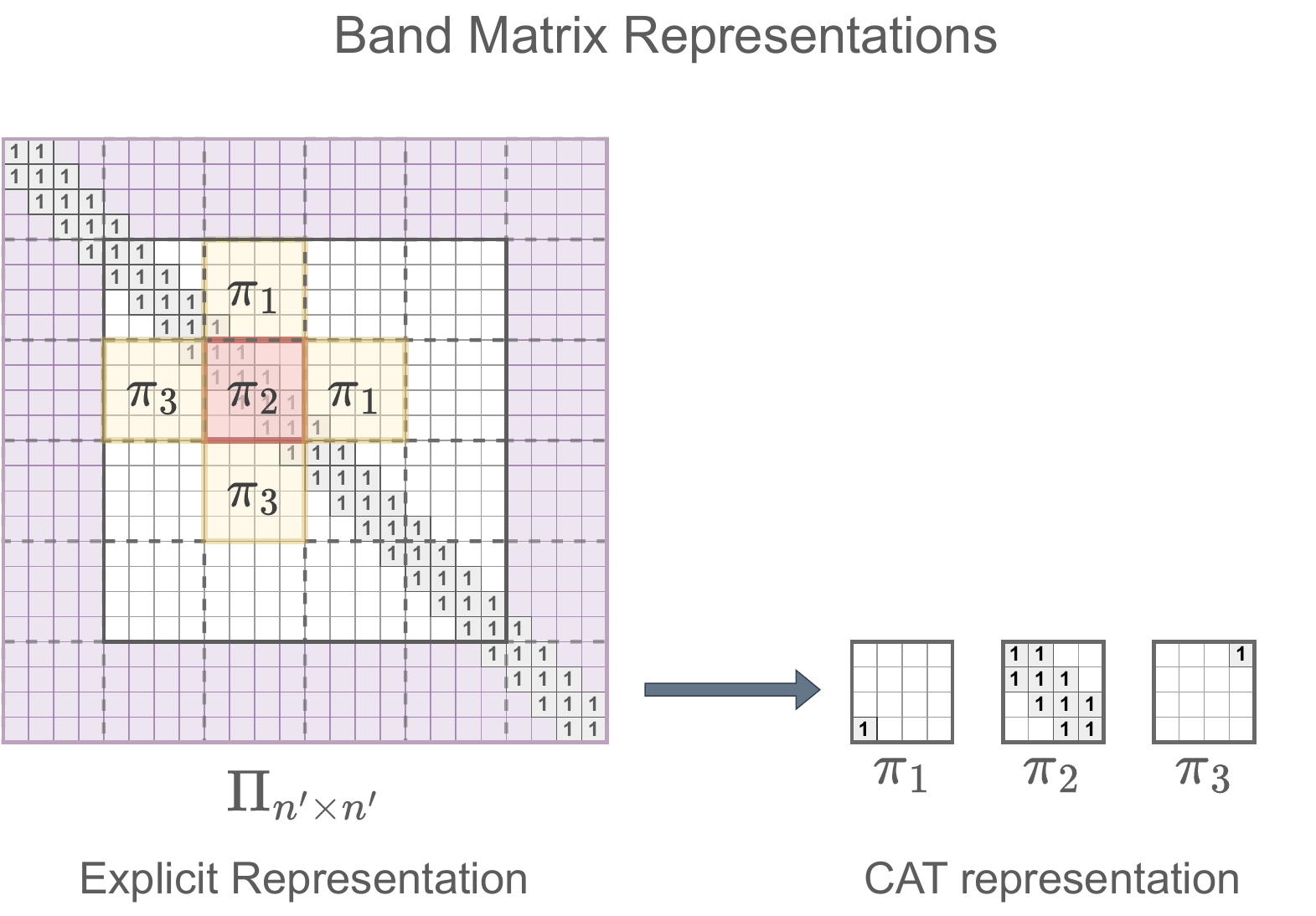}
    \caption{On the left, an explicit representation of the band matrix $\Pi$, which uses $\mathcal{O}(n^2)$ memory. On the right, the CAT representation of $\Pi$, which uses just three fragments $\pi_1, \pi_2, \pi_3$ to represent the entire matrix. Fragments are of size $4\times 4$ just for visual simplicity.}
    \label{fig:cat-band-matrix}
\end{figure}

Applying the two MMAs idea of Figure \ref{fig:cat-theory-example} directly onto each fragment one would require the \textit{in-halo} scheme, which is not efficient for large neighborhood radius as it would restrict the effective fragment area from $p \times q$ to $(p-2r)\times (q-2r)$ cells, and all fragments would have to be overlapped in $r$ cells in order to cover the entire domain properly. Moreover, an \textit{in-halo} scheme eliminates  the possibility of supporting radius values higher than the tensor core fragment size. To prevent this, CAT handles the problem differently; fragments do not restrict the effective area to their inside when $r$ increases, but instead expand to the outside allowing to simulate the entire fragment. This is achieved by adding two more MMAs for the neighbor fragments, one on each direction (horizontal and vertical). This design requires the width of the global halo of ghost cells to be a multiple of the fragment size. Considering that CAT operates with one center fragment and two adjacent ones, and the fact that the band matrix is constant, $\Pi$ can be just represented with three fragments; $\pi_1,\pi_2,\pi_3$ as shown in Figure \ref{fig:cat-band-matrix} (fragments are of size $4 \times 4$ just for visual simplicity). 

Representing $\Pi$ with just three fragments not only simplifies tensor core programming inside the GPU kernels, but it also reduces the memory usage for $\Pi$, from $\mathcal{O}(n^2)$ down to $\mathcal{O}(1)$. 

Considering that the cellular automata matrix $\Lambda$ is composed of fragments $F^\Lambda_{i,j}$, and $\Pi$ is represented by three constant fragments $\pi_1,\pi_2,\pi_3$, then the first step of CAT, \textit{i.e}, the computation of fragments $F^H_{i,j}$ in the horizontal reduction matrix $H$, is
\begin{equation}
    \label{eq:cat-h}
    F^H_{i,j} = F^\Lambda_{i,j-1} \times \pi_1 + F^\Lambda_{i,j} \times \pi_2 + F^\Lambda_{i,j+1} \times \pi_3
\end{equation}
with $i$ including the top and lower halo fragments and $j$ not including them (\textit{i.e.}, just the interior columns). All $F^H_{i,j}$ can be computed in parallel, each one handled by a different warp of threads. Once all fragments of $H$ have been computed, the second step (vertical reduction) consists of computing the fragments of matrix $R$, which are defined in terms of the fragments of $H$ as 
\begin{equation}
    \label{eq:cat-v}
    F^R_{i,j} = \pi_3 \times F^H_{i-1,j} + \pi_2 \times F^\Lambda_{i,j} + \pi_1 \times F^\Lambda_{i+1,j}
\end{equation}
this time with $i,j$ only including the interior fragments, not the halo ones.
Again, all $F^R_{i,j}$ are computed in parallel, one per warp, using the tensor cores of the GPU. It is worth mentioning that the combination of the horizontal and vertical steps and the re-use of $H$ as input for the second step allow CAT to capture the diagonal neighbors of a cell. 

\begin{figure*}[ht!]
    \centering
    \includegraphics[width=0.95\textwidth]{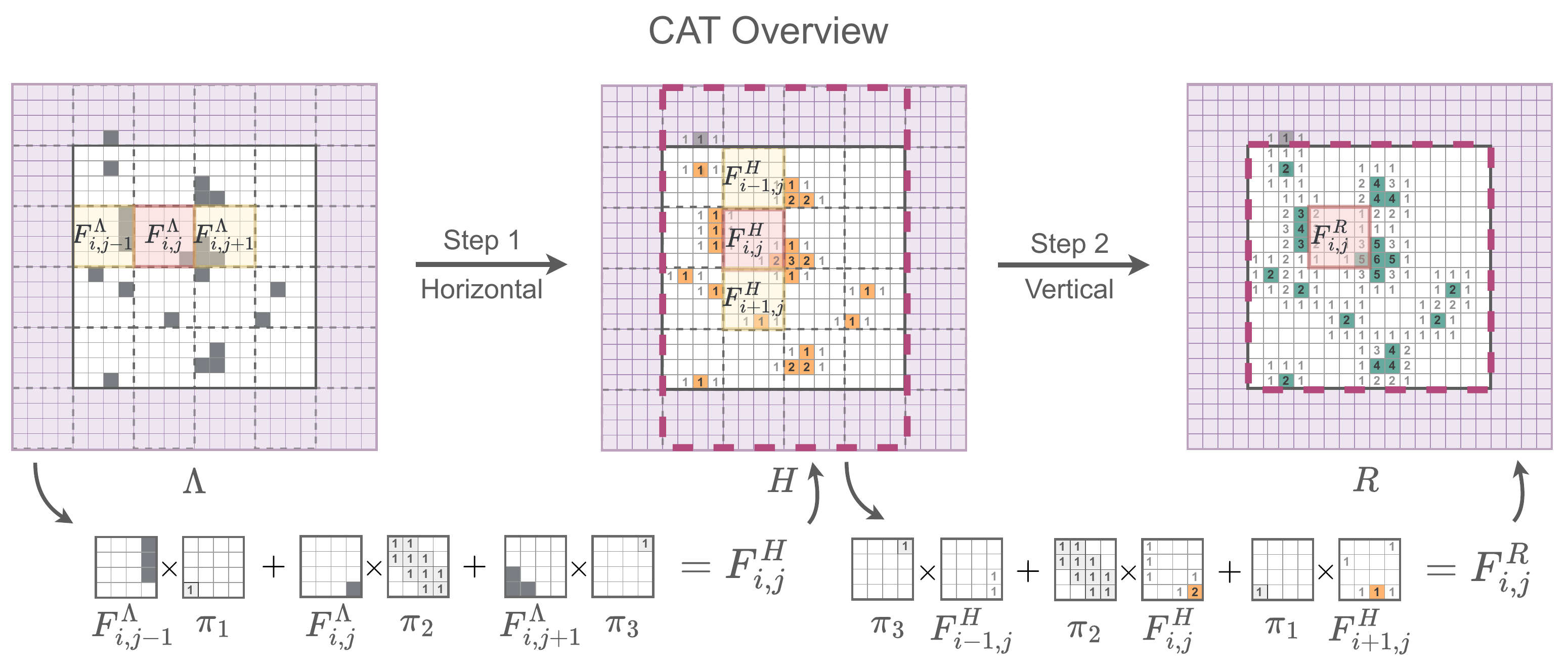}
    \caption{Overview of CAT illustrated with a Game of Life of $n\times n = 16 \times 16$ cells, neighborhood $r=1$ and periodic boundary conditions using a global halo of ghost fragments (purple). In the first step all fragments $F^H_{i,j}$ inside the dashed region of $H$ contain the horizontal reduction computed with three sequential MMAs between fragments $F^\Lambda_{i,j-1},F^\Lambda_{i,j},F^\Lambda_{i,j+1}$ and $\pi_1, \pi_2, \pi_3$. In the second step all $F^R_{i,j}$ inside the dashed region of $R$ contain the full reduction computed with three more MMAs between the fragments $\pi_3,\pi_2,\pi_1$ and $F^H_{i-1,j}, F^H_{i,j}, F^H_{i+1,j}$. This gives a total cost of six MMAs per fragment at any radius that fits in the fragment. For this example the fragments were shown as $4\times 4$ for visual clarity, but in practice the ones employed in CAT are of size $16\times 16$.}
    \label{fig:cat-overview}
\end{figure*}

The entire overview of CAT, including the horizontal/vertical steps and the regions where the fragments need to be computed (dashed regions), is illustrated in Figure \ref{fig:cat-overview} for the case of Moore neighborhood\footnote{The computational cost of CAT does not change when switching to the simplified Von Neumann neighborhood, as both neighborhoods employ the same amount of MMAs.} and fragments of $4\times 4$ for visual simplicity.
From the Figure, matrix $\Lambda$ is loaded into fragments $F^\Lambda_{i,j}$ using a per-warp logic. Then, each fragment $F^H_{i,j}$ inside the dashed region needs to be computed and is the result of computing Eq. (\ref{eq:cat-h}). Once all threads synchronize with the first step finished, they proceed to the second step which is similar, but uses $H$ as input instead of $\Lambda$ and computes Eq. (\ref{eq:cat-v}) on the dashed region of $R$. Once $R$ is computed, threads can continue in the same kernel with the transition function to each cell, using a per-thread logic. 

Matrices $\Lambda$ and $R$ need to be in GPU memory, as they play the role of \textit{in} and \textit{out}, respectively, as in any standard GPU based CA implementation. Matrix $H$ does not need to be in memory, as they are actually fragments that emerge at GPU cache level.

\subsection{CAT with large neighborhood radius}
\begin{figure}
    \centering
    \includegraphics[width=0.7\columnwidth]{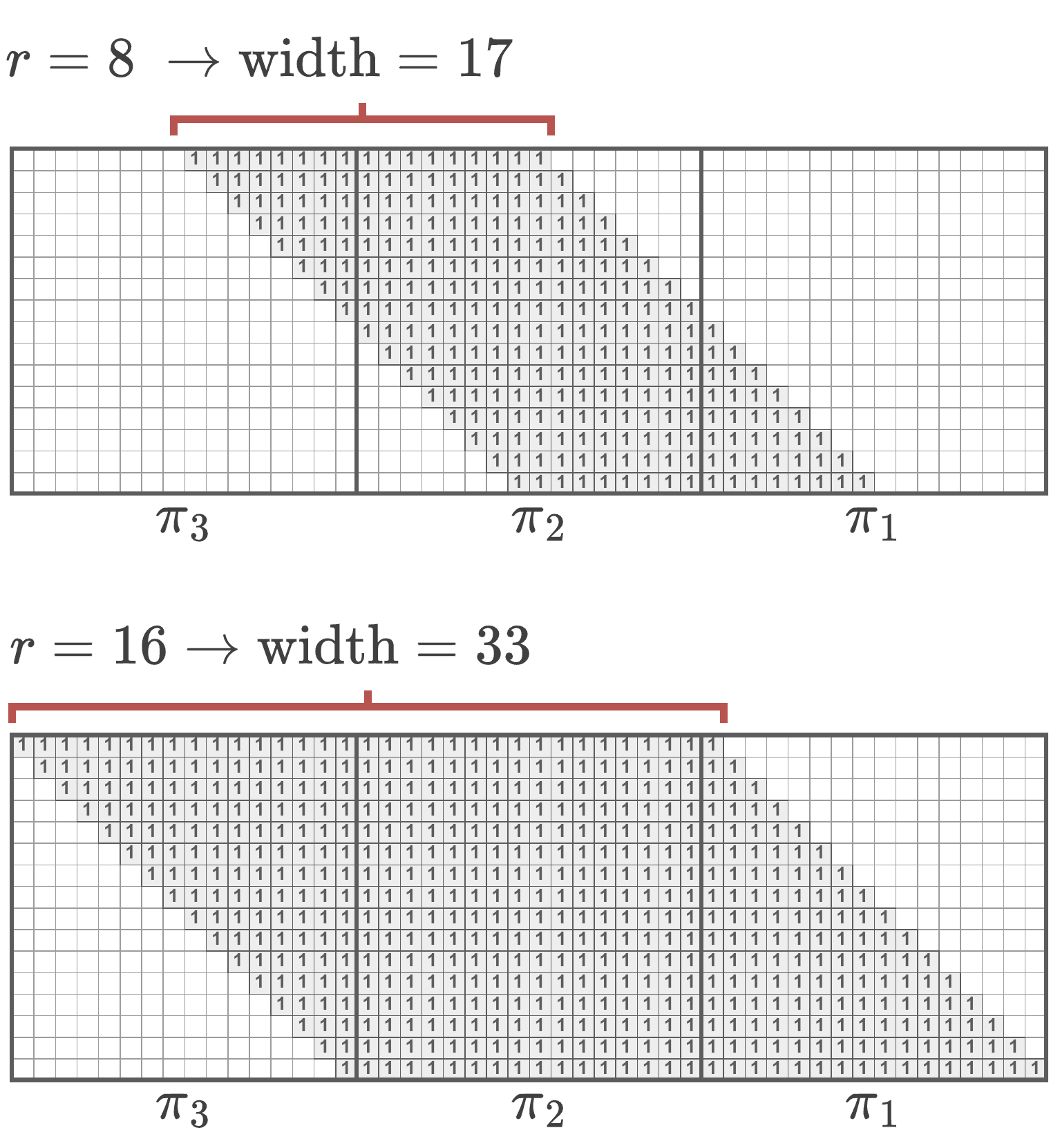}
    \caption{The band fragments at two different radius values $r=8$, $r=16$. In this case, the size of the fragments is $16 \times 16$ which is the actual size used in CAT when running in the GPU. With this fragment size, the maximum supported radius is $r=16$.}
    \label{fig:cat-radius}
\end{figure}

The main benefit of CAT is that by design it already supports simulation on large neighborhood radius up to the fragment size. This is because instead of using an \textit{in-halo} scheme, \textit{i.e.}, to restrict the effective simulation area to the inside of the fragment as $r$ increases, it uses an \textit{out-halo} scheme where the effective area is always the entire fragment and increasing $r$ only has an impact on the width of the band matrix which in practice means defining $\pi_1,\pi_2,\pi_3$ according to the $r$ value of the CA model. In other words, increasing $r$ in CAT produces the same six MMAs per fragment, but with more effective computation than with lower radius. At first, six MMAs per fragment may be seen as too much work for $r=1$, but it is greatly compensated by the fact that the cost has already been payed for higher radius, thus in theory the execution time of CAT is unaffected when increasing the radius (in the experimental results this is confirmed) as long as $r$ is within\footnote{It is possible to extend CAT to support neighborhood radiuses beyond the fragment size. This is discussed in Section \ref{sec:conclusions}.} the fragment size. In general, if the fragments are of size $p \times p$, then the maximum supported radius is $r=f$. In the case of CAT, it uses CUDA fragments of $16\times 16$ thus the maximum radius is $r=16$. Figure \ref{fig:cat-radius} shows how $\pi_1,\pi_2,\pi_3$ become for radius $r=8$ and $r=16$.

\subsection{CUDA Specific Optimizations}
CAT includes three technical optimizations: 1) Fragment-level continuous memory layout, 2) use of \textit{shared memory} and 3) optimal tile per CUDA block.

\subsubsection{Fragment-level contiguous memory layout}
\begin{figure}[ht!]
    \centering
    \includegraphics[width=0.5\columnwidth]{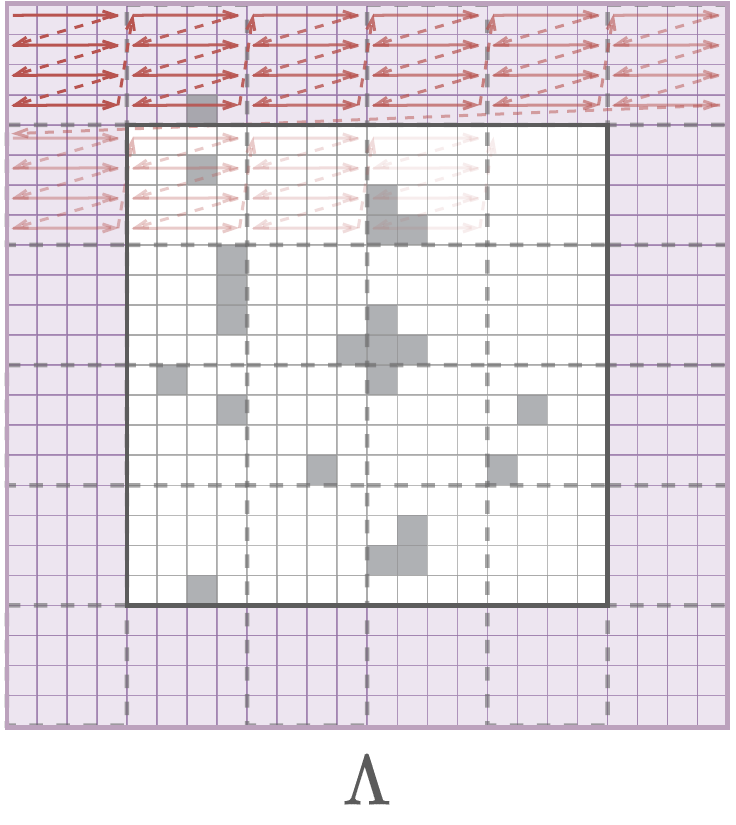}
    \caption{The fragment-level contiguous memory layout used by CAT. At the inner level, each fragment has its own row-major layout. At the outer level, the entire CA has a row-major layout of fragments.}
    \label{fig:cat-memory-layout}
\end{figure}

CAT's GPU kernel begins with each warp loading the data from GPU global memory into fragments of $16\times 16$ cells which reside at register level. Usually the CA data in global memory is in row-major layout, which is efficient for traditional GPU approaches in terms of memory accesses. However, with tensor cores, this layout is less efficient because fragments take 2D portions of memory and if the CA is in row major then there will be large memory strides of size $n$ between the rows of the fragment. This produces a slowdown in memory bandwidth, which can diminish the benefit of using CAT significantly. To overcome this issue, CAT uses a layout where each fragment is contiguous in memory. In this new layout, a group of $16\times 16 = 256$ consecutive elements corresponds to a 2D fragment. This change makes the loading of data into fragments more efficient as it is free of strides between fragment rows. Figure \ref{fig:cat-memory-layout} illustrates CAT's memory layout. 
\subsubsection{Use of shared-memory for intermediate results}
CUDA's shared memory resource was first considered for the entire process of CAT, that is, at the beginning of the kernel each CUDA block of threads would load all their corresponding cells from $\Lambda$ into a 2D \textit{shared memory} buffer, synchronize, and then perform all the remaining steps at the shared memory level until the result is written back into global memory. However, preliminary profiling of CAT showed that moving data from global to shared, and from shared to fragment registers, turned out to be slower than just loading from global to fragment. Still, CAT does use shared memory for the tile (per block) of intermediate fragments $F^H_{i,j}$, which is a fast movement of data from register level to L1 Cache \textit{shared memory} that occurs physically in the same streaming multiprocessor (SM).
Additionally, in the case of the constant fragments $\pi_1,\pi_2,\pi_3$, given that they are the same for all warps, they are generated once in shared memory. 

\subsubsection{Optimal Tile per CUDA block}
\begin{figure}[ht!]
    \centering
    \includegraphics[width=1.0\columnwidth]{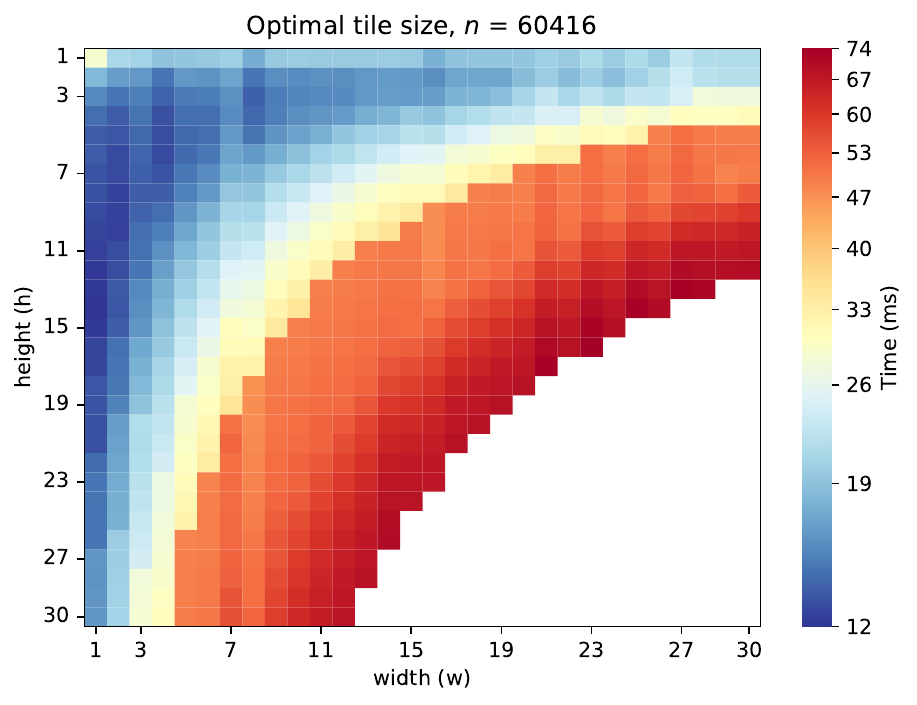}
    \caption{Heat map of the optimal shape for tile $T^H$. Long shaped tiles produce faster performance than square ones, specially narrow ones in width. The optimal tile shape for $T^H$ is near $w \times h = 1 \times 14$.}
    \label{fig:optimal-tile}
\end{figure}

With tensor cores each fragment of cells is handled by a warp of threads. One major optimization of CAT is to detach the one-to-one mapping between the CUDA block of warps and a tile of data, denoted $Q$. That is, CAT allows a tile $Q$ to have more fragments than the number of warps in its mapped CUDA block, and to have different geometries as well. This relaxation opens the possibility to explore what values of width ($w$) and height ($h$) of $Q_{w \times h}$ produces the highest performance of CAT.  
Figure \ref{fig:optimal-tile} presents a heat map where different tile shapes were explored using a large CA of size $n \times n = 60416 \times 60416$. 
From the Figure, one can note that the most efficient tile is of vertical shape, near $w \times h = 1 \times 14$. It was unexpected that a column-shaped tile would be more efficient than one with a regular shape. The experimental results shown in Section \ref{sec:experimental_results} use the optimal shape for CAT.

\subsection{Pseudocode of CAT's Kernel}
CAT uses the well known \textit{ping-pong} simulation scheme where two copies of the CA are used. One of them, the $\text{CA}_{\text{in}}$, holds the current state of cells for reading, and $\text{CA}_{\text{out}}$ is used to write the future state of cells. For each simulation step these buffers switch places. Algorithm \ref{alg:cat-kernel} presents the pseudocode of CAT's GPU kernel, with its three optimizations recently described.
\begin{algorithm}[ht!]
\caption{Pseudocode of CAT's GPU Kernel}
\label{alg:cat-kernel}
\begin{algorithmic}
	\Require $\bm{n}$:size, $\bm{\Lambda}$: In, $\bm{R}$: Out, $\bm{w},\bm{h}$:tile size, $\bm{r}$ : radius
    \State $\mathrlap{Q^H}\hphantom{S_{\pi_1}, S_{\pi_2}, S_{\pi_3}} \gets$ DeclareSharedMemoryTile$(w+2, h+2)$ 
    \State $\mathrlap{g_i,g_j}\hphantom{S_{\pi_1}, S_{\pi_2}, S_{\pi_3}} \gets$ TileGlobalOffset$(n,w,h,\text{BlockID})$
    \State $\hphantom{S_{\pi_1}, S_{\pi_2}, S_{\pi_3}}\mathllap{S_{\pi_1}, S_{\pi_2}, S_{\pi_3}} \gets$ GenSharedBandMats$(r)$   
    \State --------- SyncThreads() ---------- \Comment{\textbf{Step 1: Compute H}}
    \State $\mathrlap{W_x}\hphantom{S_{\pi_1}, S_{\pi_2}, S_{\pi_3}} \gets$ getWarpID() 
    \State $\mathrlap{W_b}\hphantom{S_{\pi_1}, S_{\pi_2}, S_{\pi_3}} \gets$ getNumWarpsPerBlock() 
    \State $\mathrlap{\pi_1,\pi_2,\pi_3}\hphantom{S_{\pi_1}, S_{\pi_2}, S_{\pi_3}}  \gets$ getBandFragments($S_{\pi_1}, S_{\pi_2}, S_{\pi_3})$
    \For{$k \gets W_x$;\ \ \ $k < w \times (h+2)$;\ \ \ $k \gets k+W_b$} 
        \State $\mathrlap{F}\hphantom{F_{i,j-1}^{\Lambda},F_{i,j}^{\Lambda},F_{i,j+1}^{\Lambda}} \gets$ ZeroFragment()
        \State $\mathrlap{(i,j)}\hphantom{F_{i,j-1}^{\Lambda},F_{i,j}^{\Lambda},F_{i,j+1}^{\Lambda}} \gets  (\lfloor k/w \rfloor, k \mod w) + (0,1)$  
        \State $F_{i,j-1}^{\Lambda},F_{i,j}^{\Lambda},F_{i,j+1}^{\Lambda} \gets$ loadFragments$(\Lambda, g_i + i, g_j + j)$ 
        \State $\mathrlap{F}\hphantom{F_{i,j-1}^{\Lambda},F_{i,j}^{\Lambda},F_{i,j+1}^{\Lambda}} \gets$ MMA$_{TC}$($F_{i,j-1}^{\Lambda}$,$\pi_1$, $[0]$ ) 
        \State $\mathrlap{F}\hphantom{F_{i,j-1}^{\Lambda},F_{i,j}^{\Lambda},F_{i,j+1}^{\Lambda}} \gets$ MMA$_{TC}$($F_{i,j}^{\Lambda}$,$\pi_2$, $F$ ) 
        \State $\mathrlap{F}\hphantom{F_{i,j-1}^{\Lambda},F_{i,j}^{\Lambda},F_{i,j+1}^{\Lambda}} \gets$ MMA$_{TC}$($F_{i,j+1}^{\Lambda}$,$\pi_3$, $F$ ) 
        \State Store($Q^H, F, i, j)$
    \EndFor
    \State --------- SyncThreads() ---------- \Comment{\textbf{Step 2: Compute R}}
    \For{$k \gets W_x$;\ \ \ $k < w \times h$;\ \ \ $k \gets k+W_b$} 
        \State $\mathrlap{F}\hphantom{F_{i,j-1}^{\Lambda},F_{i,j}^{\Lambda},F_{i,j+1}^{\Lambda}} \gets$ ZeroFragment()
        \State $\mathrlap{(i,j)}\hphantom{F_{i,j-1}^{\Lambda},F_{i,j}^{\Lambda},F_{i,j+1}^{\Lambda}} \gets  (\lfloor k/w \rfloor, k \mod w) + (1,1)$  
        \State $F_{i-1,j}^{H},F_{i,j}^{H},F_{i+1,j}^{H} \gets$ loadFragments$(Q^H, i, j)$ 
        \State $\mathrlap{F}\hphantom{F_{i,j-1}^{\Lambda},F_{i,j}^{\Lambda},F_{i,j+1}^{\Lambda}} \gets$ MMA$_{TC}$($\pi_3$, $F_{i-1,j}^{H}$, $[0]$)
        \State $\mathrlap{F}\hphantom{F_{i,j-1}^{\Lambda},F_{i,j}^{\Lambda},F_{i,j+1}^{\Lambda}} \gets$ MMA$_{TC}$($\pi_2$, $F_{i,j}^{H}$, $F$ ) 
        \State $\mathrlap{F}\hphantom{F_{i,j-1}^{\Lambda},F_{i,j}^{\Lambda},F_{i,j+1}^{\Lambda}} \gets$ MMA$_{TC}$($\pi_1$, $F_{i+1,j}^{H}$, $F$ ) 
        \State Store($\bm{R}, F, g_i + i, g_j + j)$
    \EndFor
    \State ---------------- SyncThreads() ------------------
    \For{cells assigned to current thread $\bm{t}$} \Comment{Per Thread}
        \State $c \gets $ getCell($\bm{\Lambda}, \bm{t}$)
        \State $q \gets $ getNeighborsReduction($\bm{R}, \bm{t}$)
        \State $c' \gets $ applyCARule($c, q$)
        \State Store($\bm{R}, c', \bm{t}$)
    \EndFor
\end{algorithmic}
\end{algorithm}
From the pseudocode, one can note that there are three major synchronization barriers. The algorithm starts by creating/initializing shared memory data such as the tile $\bm{Q}^H$ and shared arrays for $\bm{\pi_1}, \bm{\pi_2}, \bm{\pi_3}$ before reaching the first barrier. After this barrier, threads retrieve some warp information such as warp ID and the total number per CUDA block. Then, each warp proceeds to compute the horizontal reduction for all of its corresponding fragments using a \texttt{for} loop that has a stride of the number of warps in a CUDA block. At each iteration, the resulting horizontal reduction is stored in tile $\bm{Q}^H$. The second sync barrier ensures that all warps have finished the horizontal reduction and stored its results in $\bm{Q}^H$ before passing to the vertical reduction which is similar in logic. In general, for each reduction phase, three MMA operations are employed with the help of an auxiliary fragment $\bm{F}$ to accumulate the products and store the resulting fragment to the corresponding array; $\bm{Q}^H$ for horizontal and $\bm{R}$ for vertical. The third barrier ensures all reductions have finished, and allows threads to compute the future state of all their corresponding cells. 

\subsection{Cost Analysis of CAT}  
Computing the cost of CAT involves adding the costs of the two main steps (Figure \ref{fig:cat-overview}) as well as the application of the transition function to each cell. One way to proceed with the analysis is to obtain the parallel time of one CUDA block processing a CA tile with Algorithm \ref{alg:cat-kernel}, considering a finite number of resources (regular cores and tensor cores) residing in one GPU Streaming Multiprocessor (SM). With the parallel time per block computed, then one can expand it to the total number of tiles that need to be processed, considering that there is a finite number of SMs in a GPU chip. For this, we employ a PRAM-like model \cite{karp1988survey} in its CREW variant (Concurrent Read, Exclusive Write), but with finite resources and two extensions: i) two types of memory accesses; global memory accesses assisted by automatic L2 caching, with a cost of $C$, and cache memory (L1) with a smaller cost of $c$, \textit{i.e.}, $C = \alpha c$ with $\alpha > 1$; ii) each CUDA MMA from Eq. (\ref{eq:tensor-core-mma}) costs $\tau$ which is the actual number of one-cycle physical tensor core executions for the $p\times q \times k$ fragment. 

In CAT, the task of each CUDA block is to simulate a tile of $w \times h$ fragments from the cell matrix $\Lambda$. In the first step of Algorithm \ref{alg:cat-kernel}, the computation of each fragment $F^H_{i,j}$ involves reading three fragments from $\Lambda$ and the band fragments $\pi_1, \pi_2, \pi_3$ from shared memory, then perform three MMAs (see Figure \ref{fig:cat-overview}), and write the resulting fragment into $Q^H$. The first three memory accesses are coalesced (because of the fragment-row memory layout optimization), costing $C$ units each, while the reads on the band fragments cost $c$ units each. For the three MMAs, these cost $\tau$ units each and run one after another because one warp is in charge. Lastly, the write operation costs $c$ as it is on shared memory. With these considerations, the time for computing a fragment $F^H_{i,j}$ is
\begin{equation}
    \label{eq:cost-FH}
    \mathcal{T}_{F^H_{ij}} = 3C + 3\tau + 4c.
\end{equation}
Considering that each CUDA block gets one tile assigned, there are $w \times (h+2)$ fragments\ to compute for the tile $Q^H$. At a logical level, the number of fragments that get computed in parallel depends on the number of warps per block $W_b$, but at the physical level it depends on the number of tensor cores in a SM, which we denote $Z_{sm}$. Given that we have $Z_{sm} \le W_b$, then we can just consider $Z_{sm}$ as it is the dominant restriction. With this consideration, the parallel time for computing a tile $Q^H$ becomes:
\begin{align}
    \label{eq:cost-TH}
   \mathcal{T}_{Q^H} &= \left\lceil \frac{w\cdot (h+2)}{Z_{sm}} \right\rceil \mathcal{T}_{F^H_{i,j}}\\
                     &= \left\lceil \frac{w\cdot (h+2)}{Z_{sm}} \right\rceil (3C + 3\tau + 4c).
\end{align}

For the second step, fragments $F^R_{i,j}$ need to be computed. In this case, the warps read the three fragments from the resulting tile $Q^H$ of step 1 that lies in shared memory, thus they cost $c$ units each. The band fragments $\pi_1, \pi_2, \pi_3$ must be loaded again this time as the first operand in the MMA, each one costing $c$. Adding these costs with the three MMAs each of cost $\tau$ and with the store operation on $R$ that lies on global memory, we have that the time for computing $F^R_{i,j}$ is
\begin{equation}
    \label{eq:cost-FR}
    \mathcal{T}_{F^R_{i,j}} = 6c + 3\tau + C.
\end{equation}
In this second step, the target tile $Q^R$ has $w \times h$ fragments to compute for a given CUDA block. With this, the parallel time for computing tile $Q^R$ is
\begin{align}
    \label{eq:cost-TR}
    \mathcal{T}_{Q^R} &= \left \lceil \frac{w\cdot h}{Z_{sm}} \right \rceil \mathcal{T}_{F^R_{i,j}}\\
                    &= \left \lceil \frac{w\cdot h}{Z_{sm}} \right \rceil (6c + 3\tau + C).
\end{align}

Then, in the last stage of Algorithm \ref{alg:cat-kernel}, each thread applies the CA's transition function $f()$, of $\delta$ units, to each of its corresponding cells. For each cell in the tile, this involves reading the original cell state from $\Lambda$ ($C$ units), the neighborhood reduction from $R$ ($C$ units), then applying $f()$ ($\delta$ units) and lastly writing the result back again on $R$ ($C$ units). This gives a cost of $(\delta + 3C)$ per cell. For the parallelism at this stage, the number of regular \textit{cores} (\textit{i.e.}, the number of FP32 units as reference) are considered per SM, denoted $P_{sm}$, as it is the dominant restriction for threads in a CUDA block. With this, the parallel time for this stage of the tile is 
\begin{equation}
    \label{eq:cost-fR}
   \mathcal{T}_f = (\delta + 3C) \left \lceil \frac{(w\cdot h) (p \cdot q)}{P_{sm}} \right \rceil
\end{equation}

The last cost to include is the one at the beginning of Algorithm \ref{alg:cat-kernel}, where threads cooperatively initialize the band matrices $S_{\pi_1}, S_{\pi_2}, S_{\pi_3}$ in parallel with the regular cores $P_{sm}$ in the SM. This gives a cost per band of
\begin{equation}
    \label{eq:cost-band}
    \mathcal{T}_{S_{\pi}} = c \left \lceil \frac{p \cdot q}{P_{sm}} \right \rceil
\end{equation}
Adding the costs of Eqs. (\ref{eq:cost-TH}), (\ref{eq:cost-TR}), (\ref{eq:cost-fR}) and (\ref{eq:cost-band}), we get the following total parallel time per tile:
\begin{equation}
   \mathcal{T}_{Q_{w \times h}} = 3\mathcal{T}_{S_{\pi}} + E \cdot (\mathcal{T}_{Q^H} + \mathcal{T}_{Q^R} + \mathcal{T}_f)
\end{equation}
where $E \ge 1$ is a tile efficiency factor that models the behavior found in Figure \ref{fig:optimal-tile}, where some $w \times h$ tile shapes make CAT faster, while others make it slower. Surface $E$ was built by fitting a two-variable ($w,h$) fourth degree polynomial in the discrete points of the the heatmap. 

For a CA of $n \times n$ cells where $n$ is a multiple\footnote{This assumption is just for analysis, CAT supports any value of $n$.} of the tile size $w \times h$ and the fragment size $p \times q$, \textbf{CAT's total parallel time} running on a GPU with $\mathcal{P}$ SMs is 
\begin{equation}
    \mathcal{T}_{\text{CAT}} =\left \lceil \frac{n^2/(p \cdot q \cdot w \cdot h)}{\mathcal{P}} \right \rceil \mathcal{T}_{Q_{w\times h}}
\end{equation}

As a comparison, using the same cost model, the parallel running time of a reference (REF) baseline GPU method would do the following per cell; a thread reads its corresponding cell from global memory as well as its neighborhood of radius $r$, costing $(1+2r)^2 C$ units. Then, it computes a reduction on its neighborhood (such as addition for example), which costs $(1+2r)^2 - 1$ units and applies the transition function $f()$ which costs $\delta$ units per cell. Finally, each thread writes its resulting cell state back to global memory ($C$ units), leading to a time per cell of
\begin{equation}
    \mathcal{T}_{\text{REF}}^{\text{cell}} = (1+2r)^2 C + (1+2r)^2 -1 + \delta + C
\end{equation}
For an entire CA of $n \times n$ cells, the parallel time of a reference (REF) baseline GPU algorithm would be
\begin{align}
    \mathcal{T}_{\text{REF}} &= \left \lceil \frac{n^2}{\mathcal{P}\cdot P_{sm}} \right \rceil \mathcal{T}_{\text{REF}}^{\text{cell}}\\
    &= \left \lceil \frac{n^2}{\mathcal{P}\cdot P_{sm}} \right \rceil ((1+2r)^2 C + (1+2r)^2 -1 + \delta + C).
\end{align}

When considering the speedup of $\mathcal{T}_{\text{CAT}}$ with respect to $\mathcal{T}_{\text{REF}}$, we obtain
\begin{align}
    \label{eq:cat-speedup}
    S_{\text{CAT}} &=\frac{\mathcal{T}_{\text{REF}}}{\mathcal{T}_{\text{CAT}}} = \frac{\left \lceil \frac{n^2}{\mathcal{P}\cdot P_{sm}} \right \rceil \mathcal{T}_{\text{REF}}^{\text{cell}}}{\left \lceil \frac{n^2/(p \cdot q \cdot w \cdot h)}{\mathcal{P}} \right \rceil \mathcal{T}_{Q_{w\times h}}}.
\end{align}
From Eq. (\ref{eq:cat-speedup}) one can note that the number of streaming multiprocessors $\mathcal{P}$ is not as relevant as the internal parallel times at each one of them. To further analyze the speedup, $S_{\text{CAT}}$ must be evaluated with actual values. For this, we consider a large CA approaching the limit $n \mapsto \infty$, CAT set with a tile size of $\bm{w \times h} = 1 \times 14$ and the hardware related parameters aligned to a state of the art GPU such as NVIDIA's full GH100 chip \cite{hopperGPU}. Table \ref{tab:parameters-cost-model} summarizes these parameters. The $C$ value was defined as $6c$, and not orders of magnitude higher than $c$, because it considers the existence of the L2 cache that automatically assists the global memory on each access.
\begin{table}[ht!]
\caption{Chosen Parameters for Cost Model}
\label{tab:parameters-cost-model}
\resizebox{\columnwidth}{!}{
\begin{tabular}{|r|l|l|}
\hline
\textbf{Parameter}  & \textbf{Value}            & \textbf{Description}   \\ \hline
$\bm{n \times n}$   & limit $n \mapsto \infty$  & Representing a very large CA size.               \\ \hline
$\bm{w \times h}$   & $1 \times 14$             & CAT optimal tile size. \\ \hline
$\bm{C}$            & $6c$                      & Global + automatic L2 as a factor of $c$.  \\ \hline
$\bm{c}$            & $1$                       & Manual cache (L1) cost. \\ \hline
$\bm{p \times q}$   & $16 \times 16$            & MMA Fragment size.                    \\ \hline
$\bm{\tau}$         & $16$                      & $\#$ of internal 1-cycle calls per MMA.             \\ \hline
$\bm{P_{sm}}$       & $128$                     & Regular cores per SM.             \\ \hline
$\bm{Z_{sm}}$       & $4$                       & Tensor cores per SM.              \\ \hline
$\bm{\delta}$       & $20$                      & Cost of CA transition function $f()$.   \\ \hline
\end{tabular}
}
\end{table}

Using these parameters, Table \ref{tab:cat-scenarios} presents theoretical speedups of CAT with respect to the reference GPU approach REF, at radiuses $\bm{r}=1,4,8,16$ under different scenarios including hypothetical ones. 

% \usepackage{multirow}
\begin{table}[ht!]
\caption{CAT's Theoretical Speedups for a CA in the limit $n \mapsto \infty$.}
\label{tab:cat-scenarios}
\resizebox{\columnwidth}{!}{
\begin{tabular}{|l|l|r|r|r|r|}
\hline
\multicolumn{1}{|c|}{\multirow{2}{*}{\textbf{Scenario}}} & \multicolumn{1}{c|}{\multirow{2}{*}{\textbf{Parameter change}}}  & \multicolumn{4}{c|}{\textbf{Theoretical Speedup $\bm{S_{\text{CAT}}}$}}      \\ \cline{3-6} 
                &                                             & $r=1$         & $r=4$        & $r=8$            & $r=16$      \\ \hline
GH100 Chip      & n/a                                         & $1.20\times$  & $8.07\times$   & $27.9\times$  & $104\times$ \\\hline
More TC Units   & $\bm{Z_{sm}}: 4 \rightarrow 16$             & $1.59\times$  & $10.6\times$   & $37.1\times$    & $138\times$ \\\hline
Faster TC Units & $\bm{\tau}\ \ \ \ :  16 \rightarrow 1$      & $1.55\times$   & $10.4\times$ & $36.1\times$   & $134\times$ \\\hline
More FP Units   & $\bm{P_{sm}}:128 \rightarrow 512$           & $0.60\times$   & $4.06\times$   & $14.1\times$   & $52.5\times$ \\\hline
Regular Tiles   & $\bm{w \times h}$: $1 \times 14 \rightarrow 16 \times 16$ & $0.17\times$    & $1.15\times$    & $3.99\times$  & $14.8\times$ \\\hline
Expensive $f()$ & $\bm{\delta}$: $20 \rightarrow 10^3$        & $0.79\times$    & $1.17\times$    & $2.25\times$  & $6.44\times$ \\\hline
\end{tabular}
}
\end{table}

In the first scenario, which simulates the GH100 chip, the cost model suggests that CAT can reach significant speedups as the CA neighborhood $r$ increases, being up to $104\times$ faster than REF in a GPU chip similar to the H100. On the next scenario, which is hypothetical, increasing the number of TC units in a SM by a factor of four provides CAT a significant speedup boost of roughly $\sim 32\%$, assuming the tile size has enough fragments to keep all of these extra units active. In the next hypothetical scenario, making the TC units faster shows that it also has a strong effect, giving a boost of $\sim 28\%$ to its speedup. The third hypothetical scenario explores the case of more regular cores per SM and shows that CAT's speedup is reduced and slower than REF for $r=1$. The reason of this is because the REF implementation benefits more than CAT with this change. The fifth scenario explores the potential penalty if a non-optimal tile size is chosen for CAT, such as $w \times h = 16 \times 16$. In this case the theoretical speedup is significantly reduced to the point of being the slowest scenario for CAT at $r=1$, but manages to provide favorable speedup for the rest of the radiuses although highly penalized. The last scenario explores a more expensive transition function, showing that CAT's speedup is also penalized, becoming less favorable scenario for CAT at high radius with a maximum speedup of $6.44\times$. The reason why the amount of work in $f()$ affects CAT so much is because this work is done by the regular GPU cores, even in CAT, thus the tensor core work becomes a smaller fraction of the total work per cell. The scenarios covered in Table \ref{tab:cat-scenarios} provide useful insights on what hardware aspects and CA settings affect CAT the most. The next Section presents an experimental evaluation of CAT, comparing its performance against a reference GPU baseline (BASE) as well as with other state-of-the-art GPU techniques.

\section{Experimental Evaluation}
\label{sec:experimental_results}
\subsection{Experimental Design}
Experiments consist of measuring the time, speedup, power and energy efficiency of CAT and other state of the art GPU implementations at simulating cellular automata of $n \times n$ cells with neighborhood radius between $1 \le r \le 16$. These CA are initialized with a random uniform distribution of living states, with density $\delta$, and periodic boundary conditions. 

\subsubsection{Chosen CA tests} 
We chose instances of the Larger Than Life (LTL) family of CA \cite{griffeath1994self,evans1996larger,evans2001larger,bekaroglu2023analyzing}, which is the generalization of the well-known Game of Life (GoL) to any neighborhood radius $r$ with its own survival/birth rules. The neighborhood type is Moore, and the rule sets have been specified to produce complex phenomena for each value of $r$. The standard notation for defining LTL instances of CA is
\begin{equation}
\texttt{R}_{r}\texttt{C}_{c}\texttt{M}_{m}\texttt{S}_{s_1..s_2}\texttt{B}_{b_1..b_2}\texttt{N}_{n}
\end{equation}
where
\begin{itemize}
    \item $\texttt{R}_r$: Neighborhood radius $r$.
    \item $\texttt{C}_c$: Number of states ($c$) per cell.
    \item $\texttt{M}_m$: Include ($m=1$) or not ($m=0$) the center cell.
    \item $\texttt{S}_{s_1..s_2}$: Survival range $[s_1..s_2]$ for living cells. 
    \item $\texttt{B}_{b_1..b_2}$: Birth range $[b_1..b_2]$ for dead cells. 
    \item $\texttt{N}_n$: Moore ($n=M$) or Von Neumann ($n=N$) neighborhood.
\end{itemize}
The LTL instances used for the tests are presented in Table \ref{tab:ltl-instances}, each one using a certain value of $r$ and an initial density $\delta$ of living cells. Some of these LTL instances are known in the literature and cited, such as Bosco's rule by Evans \cite{evans2005bosco}, or Bugsmovie by Griffeath \cite{griffeath1994self}, among others. The other CA instances in the table are custom variations made by our team, with their birth and survival ranges adapted to the specific value of $r$. 
\begin{table}[ht!]
\caption{Larger than Life instances for testing at different $r$.}
\label{tab:ltl-instances}
\resizebox{\columnwidth}{!}{
\begin{tabular}{|l|l|l|l|}
\hline
\textbf{$\bm{r}$}  & \textbf{LTL Instance}                           & $\bm{\delta}$ & \textbf{Name}                                       \\ \hline
1             & \ltl{1}{2}{0}{2}{3}{3}{3}{M}            & $0.07$   & Game of Life \cite{gardner1970fantastic}   \\ \hline
2             & \ltl{2}{2}{0}{7}{12}{8}{11}{M}          & $0.15$   & Starry Night (custom)  \\ \hline
3             & \ltl{3}{2}{0}{15}{23}{14}{17}{M}        & $0.25$   & Boiling Gnocchi (custom)  \\ \hline
4             & \ltl{4}{2}{0}{40}{80}{41}{80}{M}        & $0.50$    & Majority  \cite{griffeath1994self}         \\ \hline
5             & \ltl{5}{2}{0}{35}{59}{34}{45}{M}        & $0.21$   & Bosco's Rule \cite{evans2005bosco}         \\ \hline
6             & \ltl{6}{2}{0}{49}{81}{46}{65}{M}        & $0.22$   & Radiation (custom)  \\ \hline
7             & \ltl{7}{2}{0}{101}{201}{75}{170}{M}     & $0.29$   & Waffles \cite{MirekWojtowicz}  \\ \hline
8             & \ltl{8}{2}{0}{163}{223}{74}{252}{M}     & $0.23$   & Globe \cite{MirekWojtowicz}      \\ \hline
9             & \ltl{9}{2}{0}{108}{181}{100}{140}{M}    & $0.24$   & Gravity (custom) \\ \hline
10            & \ltl{10}{2}{0}{122}{211}{123}{170}{M}   & $0.25$   & Bugsmovie  \cite{griffeath1994self}        \\ \hline
11            & \ltl{11}{2}{0}{156}{265}{147}{205}{M}   & $0.24$   & Broken Ships (custom) \\ \hline
12            & \ltl{12}{2}{0}{170}{296}{170}{240}{M}   & $0.25$   & Scaled GOL \cite{griffeath1994self} \\ \hline
13            & \ltl{13}{2}{0}{213}{364}{203}{283}{M}   & $0.25$   & The Cleansing (custom) \\ \hline
14            & \ltl{14}{2}{0}{245}{420}{234}{326}{M}   & $0.25$   & Scaled Bugsmovie \cite{griffeath1994self} \\ \hline
15            & \ltl{15}{2}{0}{170}{296}{170}{240}{M}   & $0.28$   & Pretzels \cite{griffeath1994self} \\ \hline
16            & \ltl{16}{2}{0}{170}{296}{170}{300}{M}   & $0.26$   & Tangy Ramen (Custom) \\ \hline
\end{tabular}
}
\end{table}
It is worth clarifying that choosing one or another LTL instance does not affect the performance of the GPU approaches, as they are only affected by $n$ and $r$. 

\subsubsection{Approaches Selected for Evaluation}
Table \ref{tab:approaches} lists the GPU implementations that were selected for performance evaluation, with their source codes available at \href{https://github.com/temporal-hpc/CAT}{https://github.com/temporal-hpc/CAT}.
\begin{table}[ht!]
\caption{GPU Approaches selected for evaluation}
\label{tab:approaches}
\resizebox{\columnwidth}{!}{
\begin{tabular}{|l|l|}
\hline
\textbf{Approach} & \textbf{Main idea and GPU implementation details}   \\ \hline
BASE                &   \begin{tabular}[c]{@{}l@{}}Baseline global memory approach.\\
                        \textbullet\ Implemented by our team.\\
                        \textbullet\ Each thread simulates one cell (\texttt{char}) in global memory.\\
                        \textbullet\ CUDA Block Size: $B_x \times B_y = 32 \times 32$ threads.\\
                        \textbullet\ Block Size: $B_x \times B_y = 32 \times 32$ threads.\\
                        \textbullet\ Neighborhood radius: $r \in [1..16]$.\\
                        \end{tabular}\\
                        \hline
SHARED                & \begin{tabular}[c]{@{}l@{}}Classic Shared Memory approach \cite{topa2013using}.\\
                        \textbullet\ Implemented by Millan et al. \cite{millan2017performance} supporting $r \in [1..5]$.\\
                        \textbullet\ \textit{Out-halo}: sh-mem of $(B_x + 2r) \times (B_y + 2r)$ cells.\\
                        \textbullet\ One cell (\texttt{char}) per thread.\\
                        \textbullet\ CUDA Block Size: $B_x \times B_y = 32 \times 32$ threads.\\
                        \textbullet\ \textbf{[us]} Radius extended to $r \in [1..16]$.\\
                        \end{tabular}\\
                        \hline
\begin{tabular}[c]{@{}l@{}} \textbf{CAT}\\(proposed)\end{tabular} & \begin{tabular}[c]{@{}l@{}}Proposed Tensor Core (TC) based approach.\\
                         \textbullet\ Implemented by our team.\\
                         \textbullet\ Neighborhood reduction through Tensor Core MMAs.\\
                         \textbullet\ Multiple TC fragments (\texttt{FP16}) per warp.\\
                         \textbullet\ \textit{Out-halo}: adjacent fragments for boundary cells.\\
                         \textbullet\ Uses sh-mem for intermediate results.\\
                         \textbullet\ CUDA Block Size: $B_x \times B_y = 16 \times 16$ threads.\\
                         \textbullet\ Neighborhood radius: $r \in [1..16]$.
                         \end{tabular}\\
                         \hline
COARSE                & \begin{tabular}[c]{@{}l@{}}Alternative shared memory approach.\\
                        \textbullet\ Implemented by our team.\\
                        \textbullet\ Thread Coarsening \cite{coarsening10.1145/2503210.2503268} $\rightarrow$ Multiple cells (\texttt{char}) per thread.\\
                        \textbullet\ \textit{Out-halo}: sh-mem of $(80 + 2r) \times (80 + 2r)$ cells.\\
                        \textbullet\ Neighborhood radius: $r \in [1..16]$.
                        \end{tabular}\\ 
                        \hline
MCELL                 & \begin{tabular}[c]{@{}l@{}}Multi-cell + shared memory approach.\\
                        \textbullet\ Original code by Millan et al. \cite{millan2017performance}, supporting $r \in [1..5]$.\\
                        \textbullet\ Two adjacent cells (\texttt{char}) per thread for register re-use.\\
                        \textbullet\ CUDA Block Size: $B_x \times B_y = 16 \times 16$ threads.\\
                        \textbullet\ \textbf{[us]} \textit{Out-halo}: sh-mem of $(2B_x + 2r) \times (B_y + 2r)$ cells.\\
                        \textbullet\ \textbf{[us]} Improved memory access on two-cell neighborhood.\\
                        \textbullet\ \textbf{[us]} Radius extended to $r \in [1..16]$.\\
                        \end{tabular}\\ 
                        \hline
PACK                  & \begin{tabular}[c]{@{}l@{}}Packet coding technique.\\
                        \textbullet\ Idea and code by Cagigas et al. \cite{cagigas2022efficient} supporting $r=1$.\\
                        \textbullet\ 64-bit words codified as eight \texttt{char} (8-bit) cells.\\
                        \textbullet\ No use of shared memory $\rightarrow$ faster in global memory.\\
                        \textbullet\ CUDA Block Size: $B_x \times B_y = 16 \times 16$ threads.\\
                        \textbullet\ \textbf{[us]} Radius extended to $r \in [1..16]$.\\
                        \end{tabular} \\ \hline
\end{tabular}
}
\end{table}

BASE, CAT and COARSE were implemented by our team using CUDA C/C++. BASE is a GPU baseline and corresponds to a standard GPU implementations of stencil computation \cite{d2012cellular} where each thread simulates one cell reading its entire neighborhood directly from global memory. COARSE is a shared memory approach with an \textit{out-halo} design that also includes thread coarsening \cite{coarsening10.1145/2503210.2503268}. It uses a large shared memory tile of $(80 + 2r)\times (80 + 2r)$ cells per block, and each thread simulates multiple cells with a stride of the block size. 

SHARED, MCELL and PACK correspond to related works that made their implementation available using CUDA C/C++ as well. We extended these implementations to support $1 \le r \le 16$ and also optimized some of them. For SHARED, the core logic of the technique was extended to $r \in [1..16]$ with the expected code changes towards radius generalization. For MCELL the case was less straightforward, extending it to $r \in [1..16]$ required changing the design from \textit{in-halo} to \textit{out-halo} in order to run efficient in the range of $r$, and also required improving the memory access such that the two-cell neighborhood is read once for each thread at any $r$. In the case of PACK, the extension of Cagigas et al. implementation \cite{cagigas2022efficient} to $r \in [1..16]$ followed their scheme but now considering that when $8 < r \le 16$ the horizontal neighbor cells are found in the two consecutive 64-bit words to the left/right, and not one as in the original case. It is worth mentioning that these extended implementations run as fast, or faster than before when executed using their original supported radiuses. More details on the core idea of the last three approaches can be found in Section \ref{sec:related_work}.  

In terms of precision, CAT uses \texttt{FP16} types for the neighborhood reduction, as it is currently the most suitable type offered by GPU tensor cores in which square fragments can be defined (\textit{i.e.}, full \texttt{INT32} MMAs are still not supported in square shape). Still, converting the floating point reduction into \texttt{INT32} provided correct simulations throughout the range $1 \le r \le 16$. The reason why this works is because \texttt{FP16} has a 10 bit mantissa, which corresponds to a precision of $10+1$, then $2^{11}$ is the largest integer that can be precisely represented, which is sufficient to store the maximum possible amount of living neighbors with radius $r=16$ (\textit{i.e.} $33\times33-1$ cells). The rest of the approaches employ \texttt{INT32} precision for neighbor counting and \texttt{char} precision for the cell state.

\subsubsection{Performance measures}
The performance measures are \textit{time} in milliseconds and \textit{speedup} with respect to the BASE approach. Each measure is taken as the average of multiple realizations, varying $n$ in the range $n \in [1024, 60416]$. For the lower values of $n$ we used up to 32 realizations while for high $n$ we used up to four. Each realization measures the average running time for simulating the given LTL instance, at a given $n$, for 25 time steps. These measuring settings ensured averages with a standard error of $1\%$ or less. 

The energy measurements were done by running a large size simulation of $n \times n = 60416 \times 60416$ cells and measuring the instant GPU Power in Watts (W) as a time series for 1000 simulation steps, using the \texttt{nvml} library from Nvidia. From these GPU power time series, the average total energy per simulation step is obtained in Joules (J), and the average energy efficiency per simulation step is obtained as $Cells/J$.

\subsubsection{Testing Platform}
Table \ref{tab:testing-platform} presents the testing platform. 
\begin{table}[ht!]
    \caption{Testing Platform}
    \label{tab:testing-platform}
    \centering
    \begin{tabular}{|c|l|}
    \hline
    \textbf{Property} & \textbf{Value} \\ \hline 
    OS          & Linux Ubuntu 22.04 LTS \\ \hline
    CPU         & 2$\times$ 52-core Intel Xeon Platinum 8470 \\ \hline
    RAM         & 128GB RAM \\ \hline
    GPU         & NVIDIA H100 SXM5 80GB HBM3 \\ \hline
    \end{tabular}
\end{table}
\subsection{Performance Results}
\begin{figure*}[ht!]
    \centering
    \includegraphics[width=0.24\textwidth]{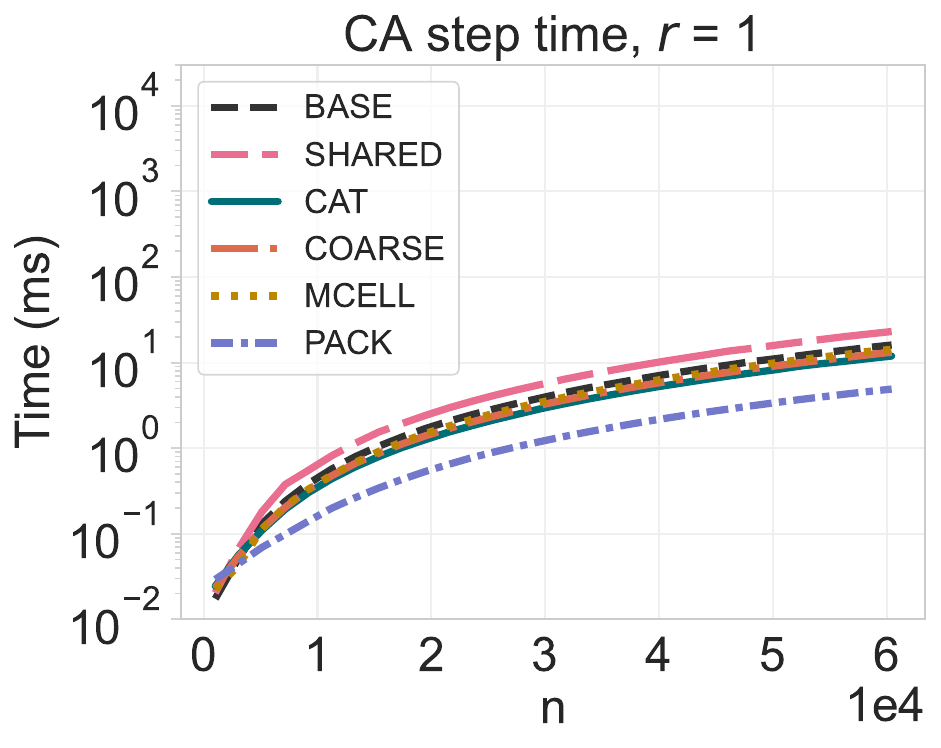}
    \includegraphics[width=0.24\textwidth]{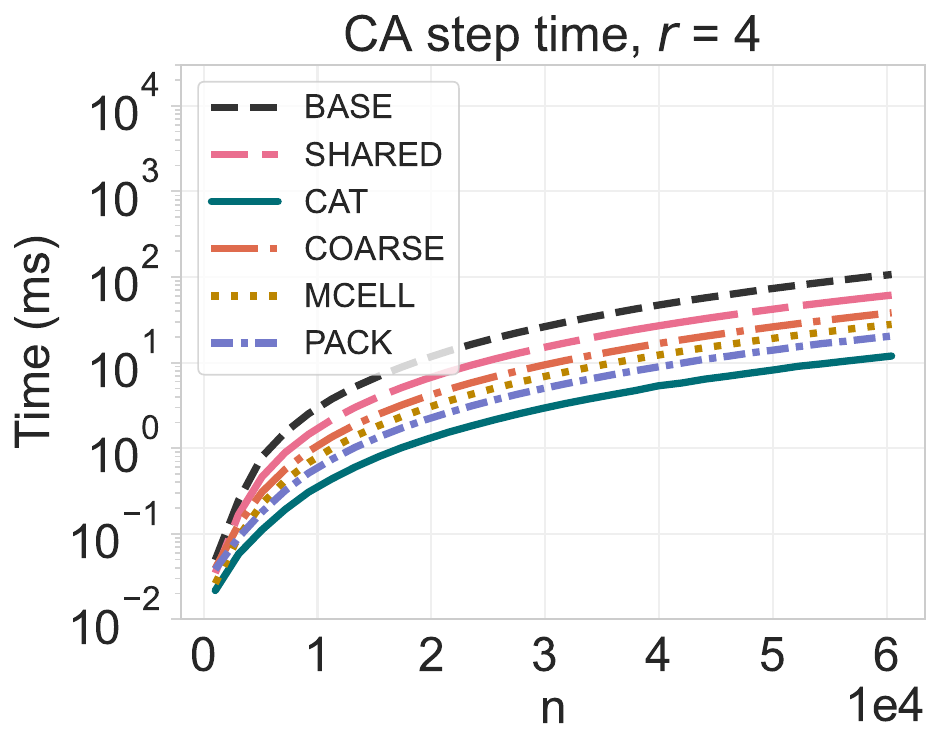}
    \includegraphics[width=0.24\textwidth]{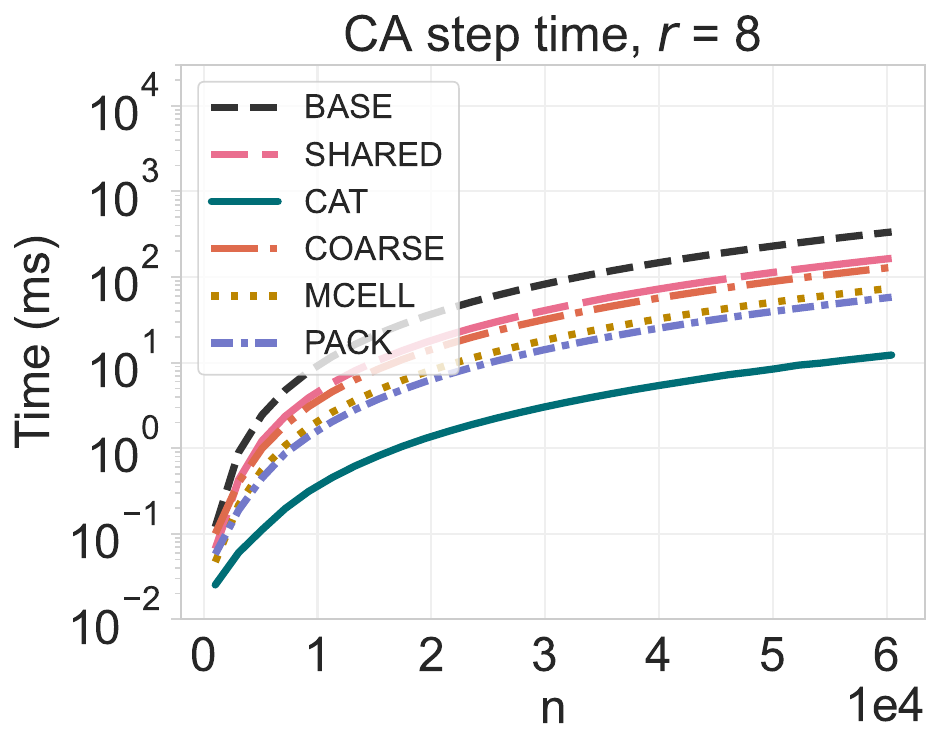}
    \includegraphics[width=0.24\textwidth]{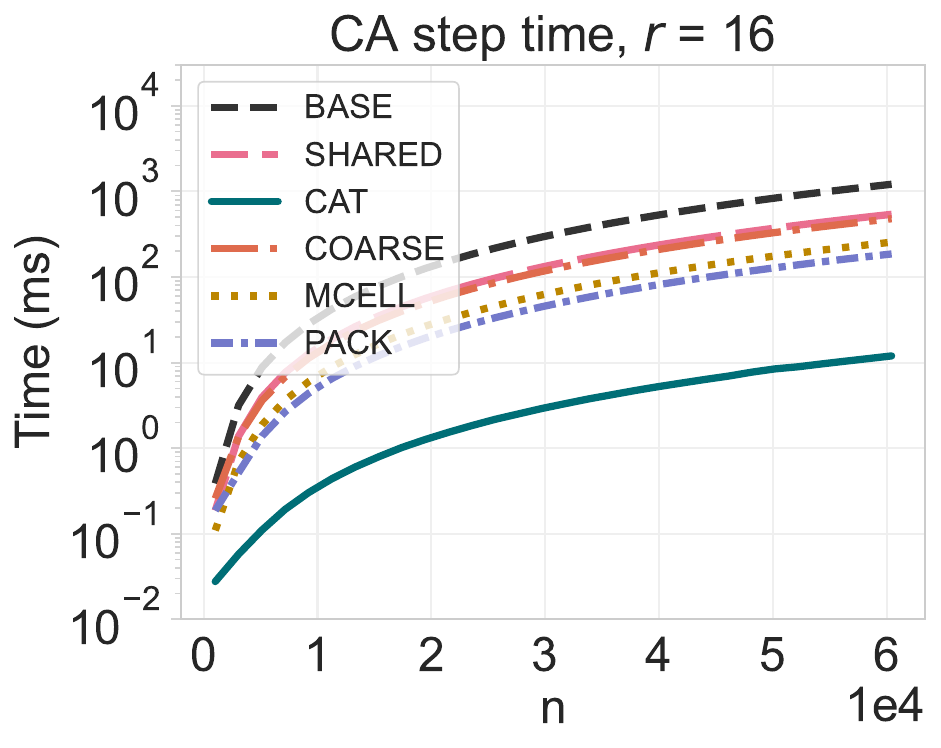}
    \includegraphics[width=0.24\textwidth]{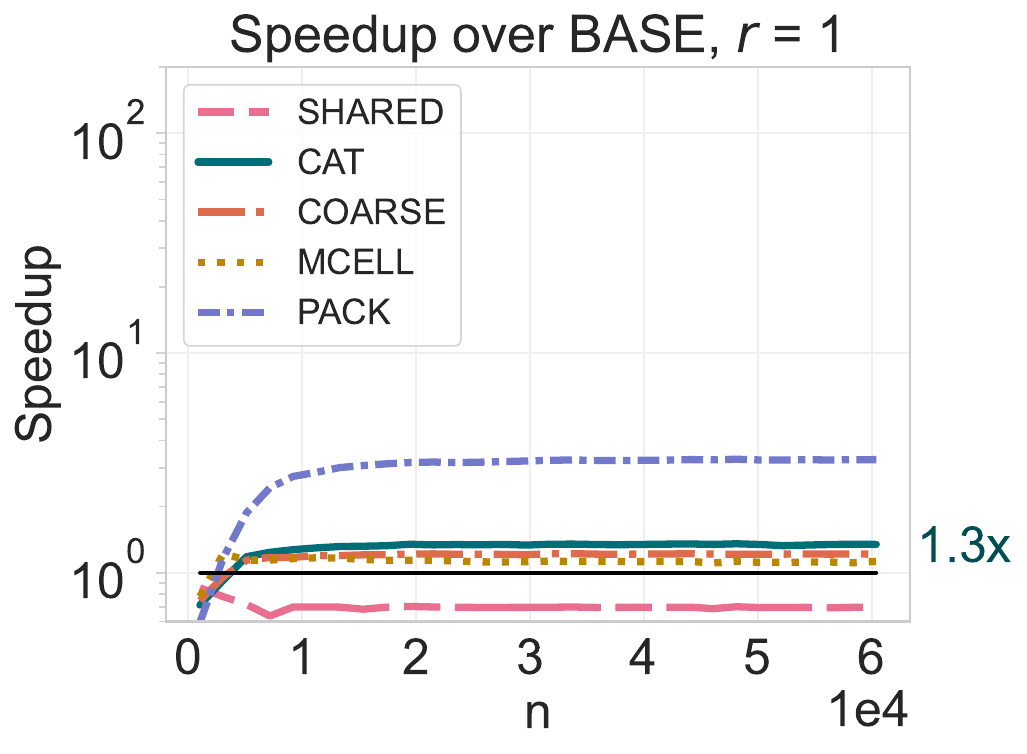}
    \includegraphics[width=0.24\textwidth]{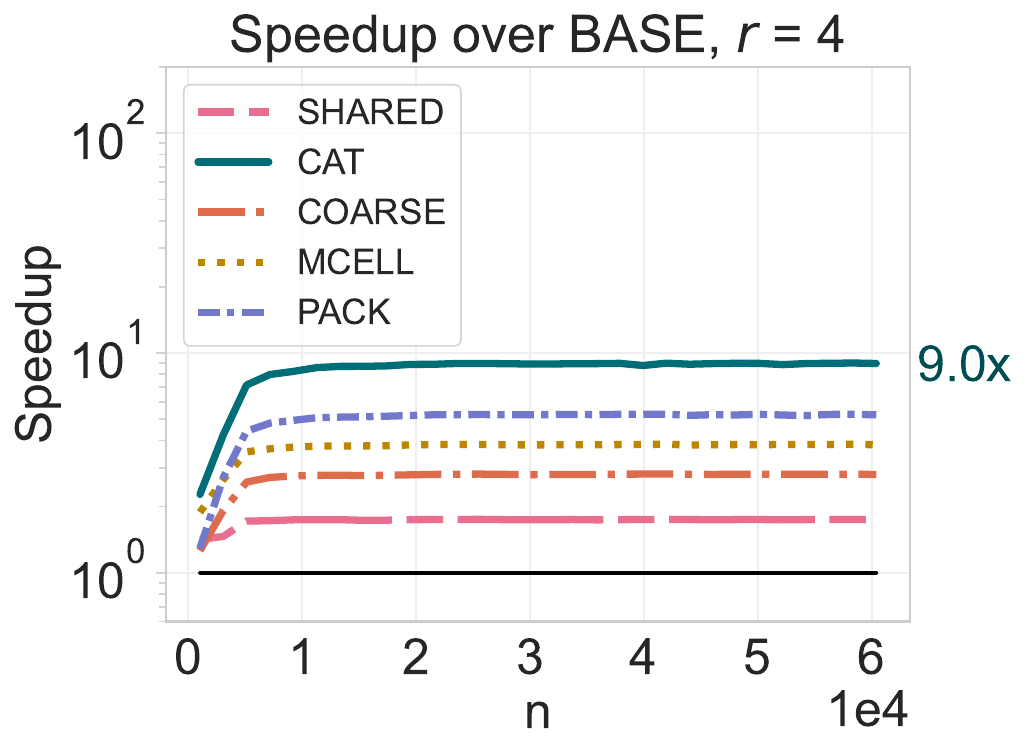}
    \includegraphics[width=0.24\textwidth]{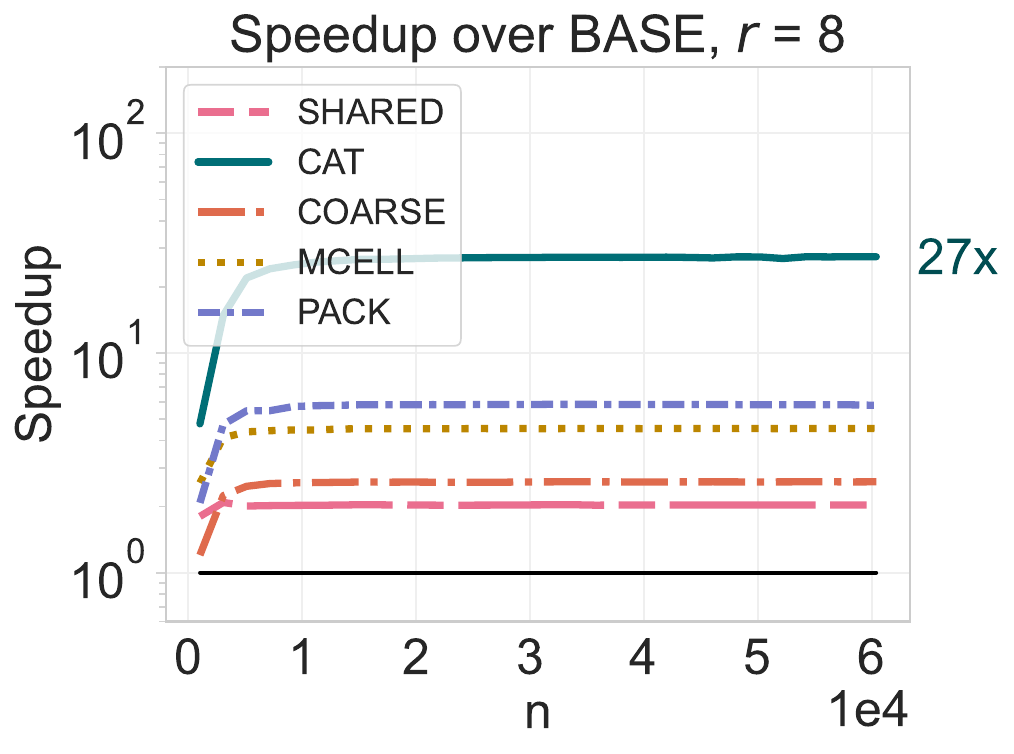}
    \includegraphics[width=0.24\textwidth]{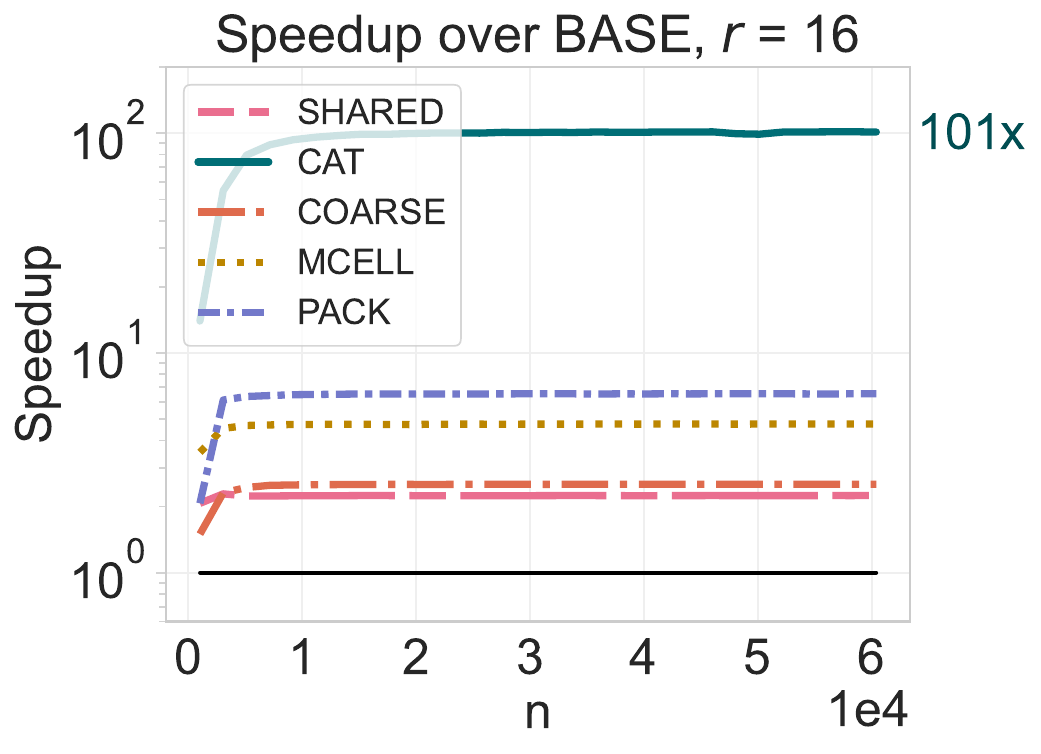}
    \caption{Time and Speedup for all approaches at different values of neighborhood $r$.}
    \label{fig:time-speedup}
\end{figure*}

Figure \ref{fig:time-speedup} presents the time and speedup of the selected approaches. The first row shows the average time per simulation for radiuses $r=1,4,8,16$, while the second row shows the speedups with respect to BASE. The results show that CAT keeps its running time constant as $r$ increases while the other approaches take more time to complete. This makes CAT's speedup to increase with $r$, starting as the second fastest at $r=1$ with a speedup of $1.3\times$, but becoming the fastest one at $r=4, 8, 16$ with speedups of $9\times$, $27\times$, and $101\times$ respectively. These results also agree with the cost model and its theoretical speedups from Table \ref{tab:cat-scenarios} when the H100 was assumed. For the other approaches, one can observe that their speedups cluster into three groups as $r$ increases; the first top group with just CAT, the second group with PACK followed by MCELL, and the third group with COARSE followed by SHARED. It is worth noting that PACK is the fastest approach at $r=1$ with $\sim 3\times$ of speedup, which translates to being $2.3\times$ faster than CAT. On the other hand, at $r=16$ CAT is $\sim 14\times$ faster than PACK. Another behavior to note is that COARSE performs better than SHARED as a shared memory based solution, and it avoids being slower than the BASE at $r=1$ as SHARED did, which is a known issue of shared memory solutions in modern GPUs \cite{schafer2011high,RENC2022101538}. This difference between COARSE and SHARED has less of an impact at higher $r$, where both a much closer in performance.

\begin{figure}[ht!]
    \centering
    \includegraphics[width=0.24\textwidth]{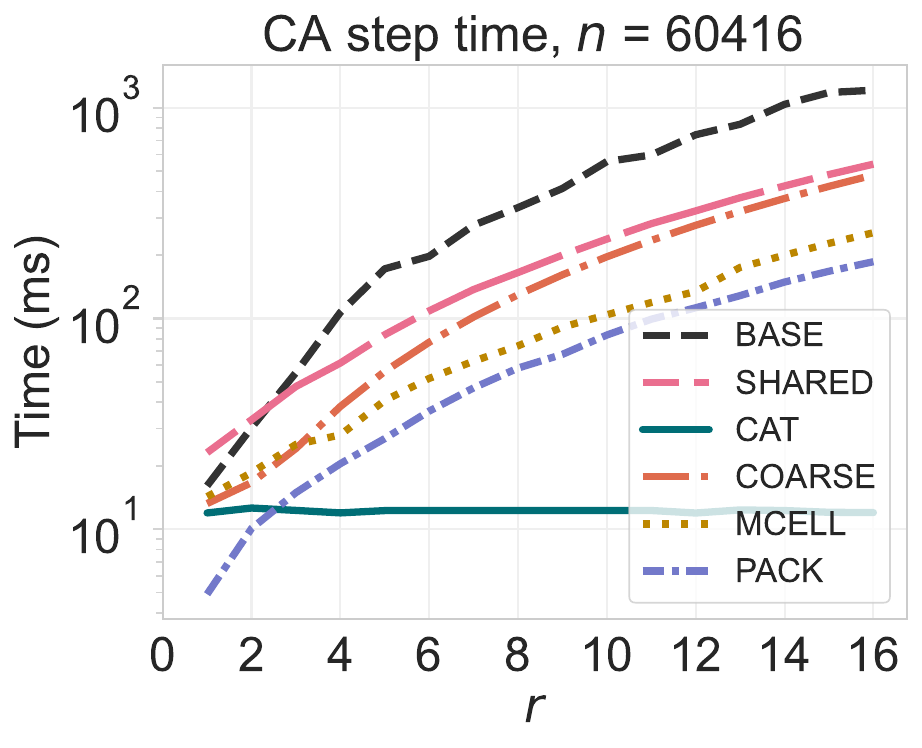}
    \includegraphics[width=0.24\textwidth]{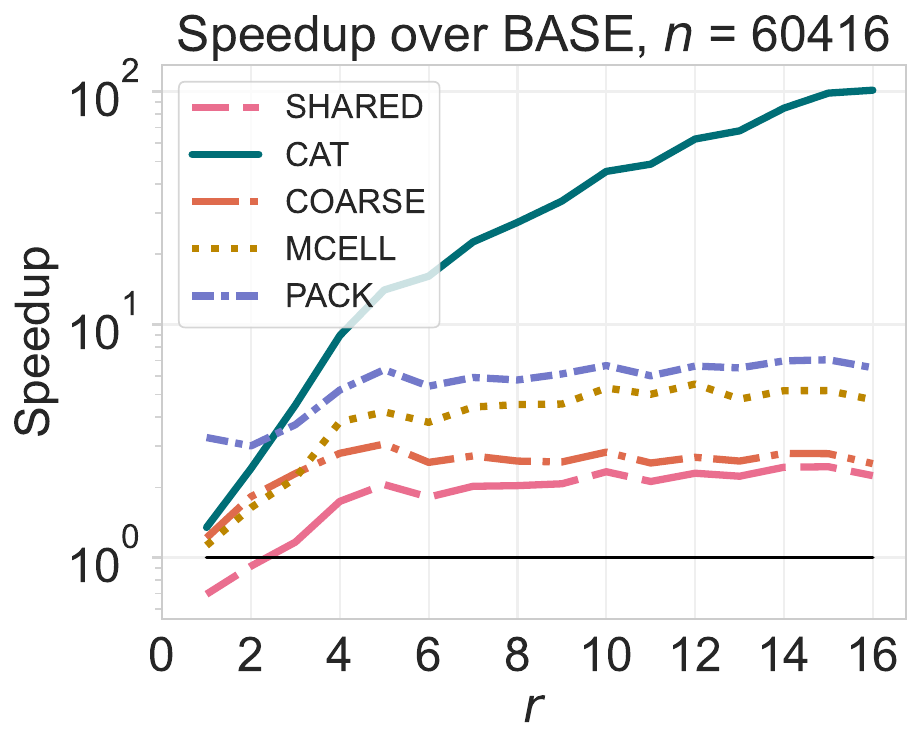}
    \caption{Impact of the neighborhood radius $r$ in the time and speedup of all approaches for a large CA of $n \times n = 60416 \times 60416$ cells.}
    \label{fig:perf-radius}
\end{figure}

More details are given in Figure \ref{fig:perf-radius}, which shows the performance impact from increasing the neighborhood radius $r$, given a large CA of $n \times n = 60416 \times 60416$ cells. 

The plots show that $r=3$ is the crossover point where CAT surpasses PACK and becomes the fastest approach for the remaining range of $r$. Coincidently, it is also the crossover point where MCELL surpasses COARSE, and where SHARED surpasses BASE. Another behavior to note is that CAT's speedup increases with $r$ being up to two orders of magnitude faster than BASE, whereas the other approaches converge at specific values, with PACK being the second fastest approach reaching near $7\times$ of speedup.  

\subsection{Energy Efficiency}
\begin{figure*}[ht!]
    \centering
    \includegraphics[width=0.24\textwidth]{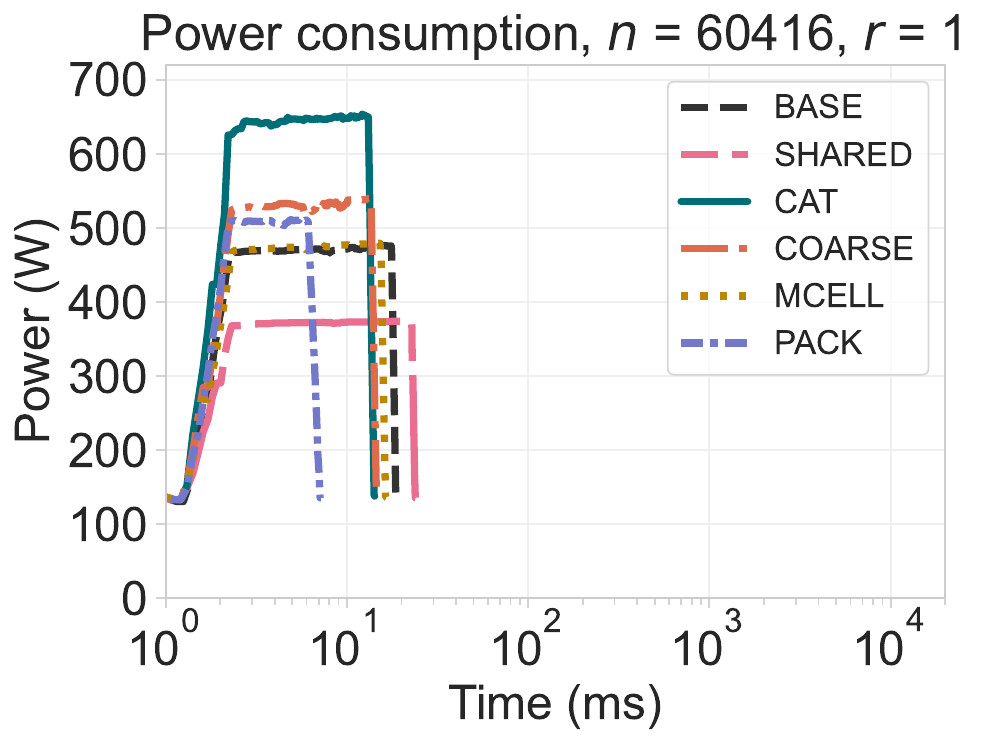}
    \includegraphics[width=0.24\textwidth]{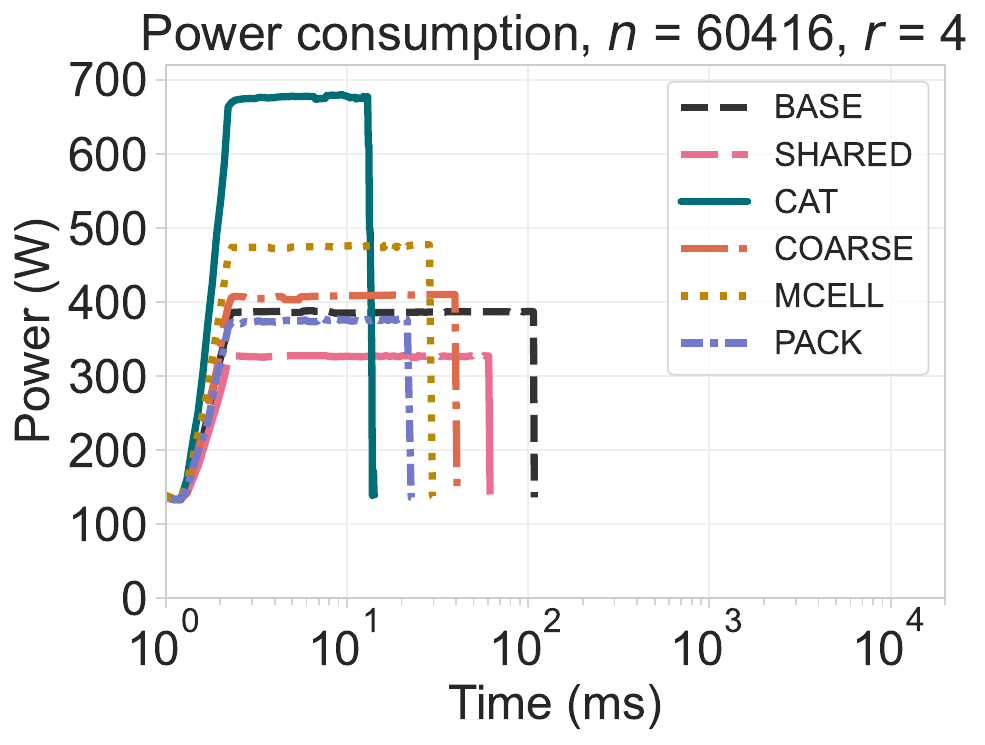}
    \includegraphics[width=0.24\textwidth]{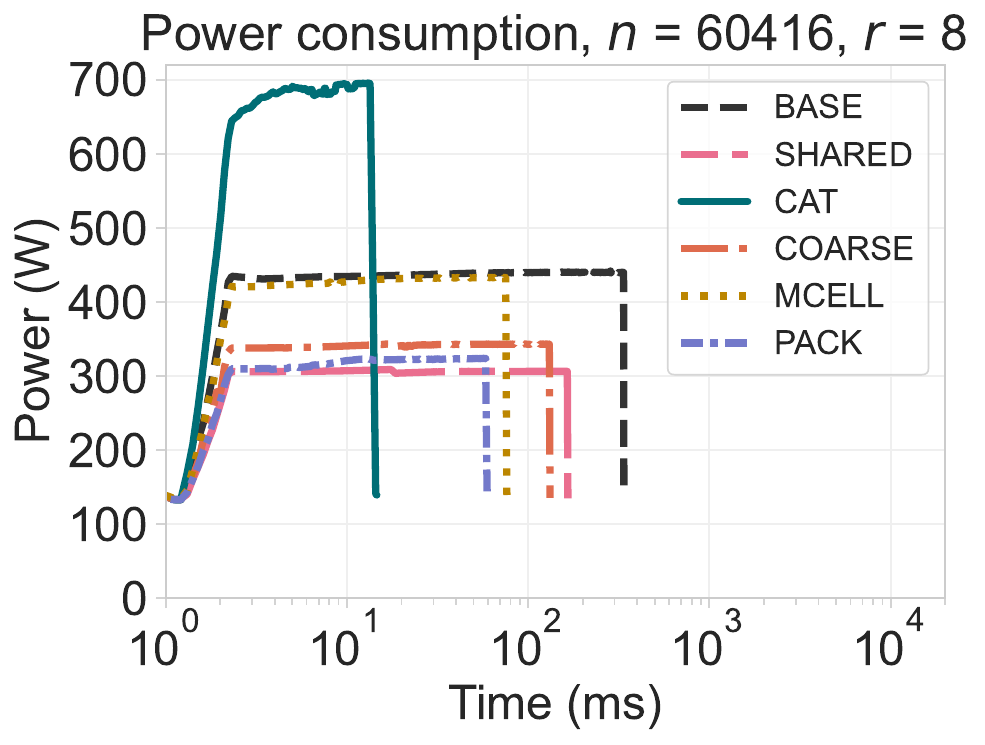}
    \includegraphics[width=0.24\textwidth]{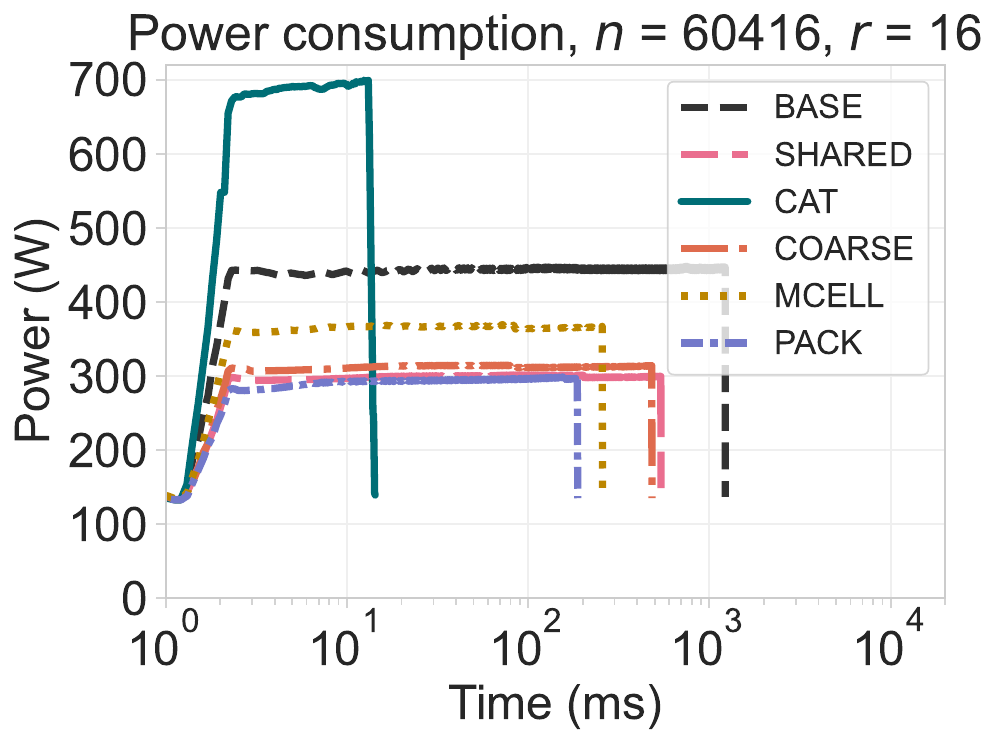}
    \includegraphics[width=0.24\textwidth]{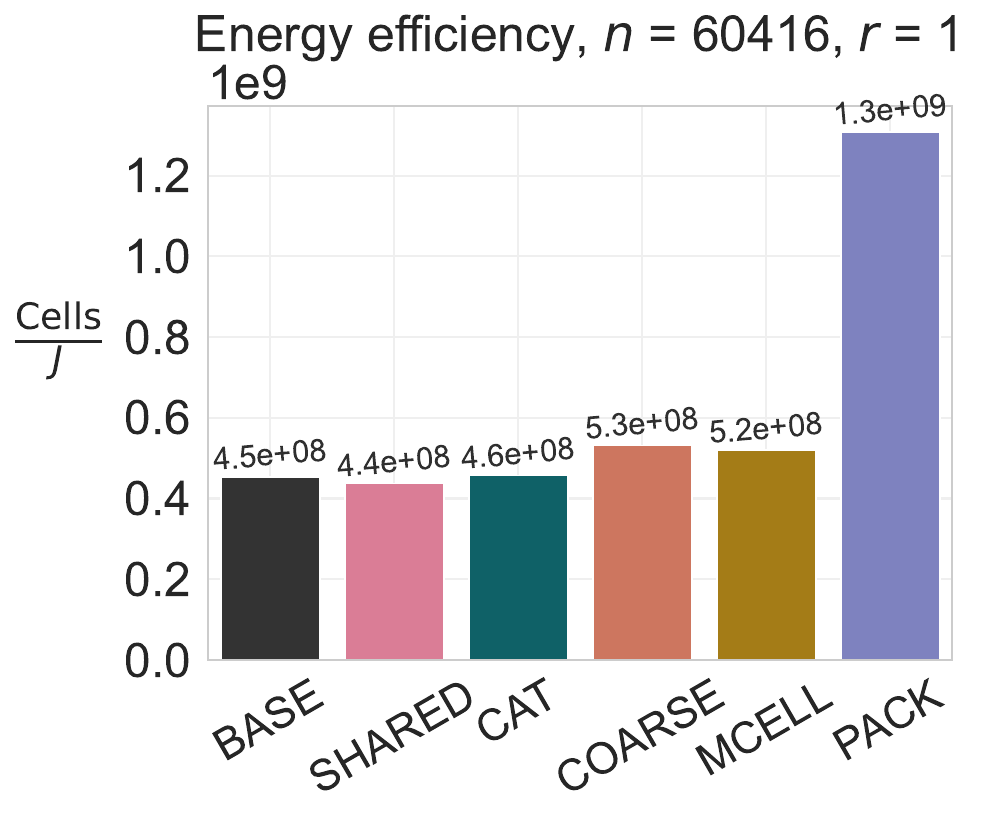}
    \includegraphics[width=0.24\textwidth]{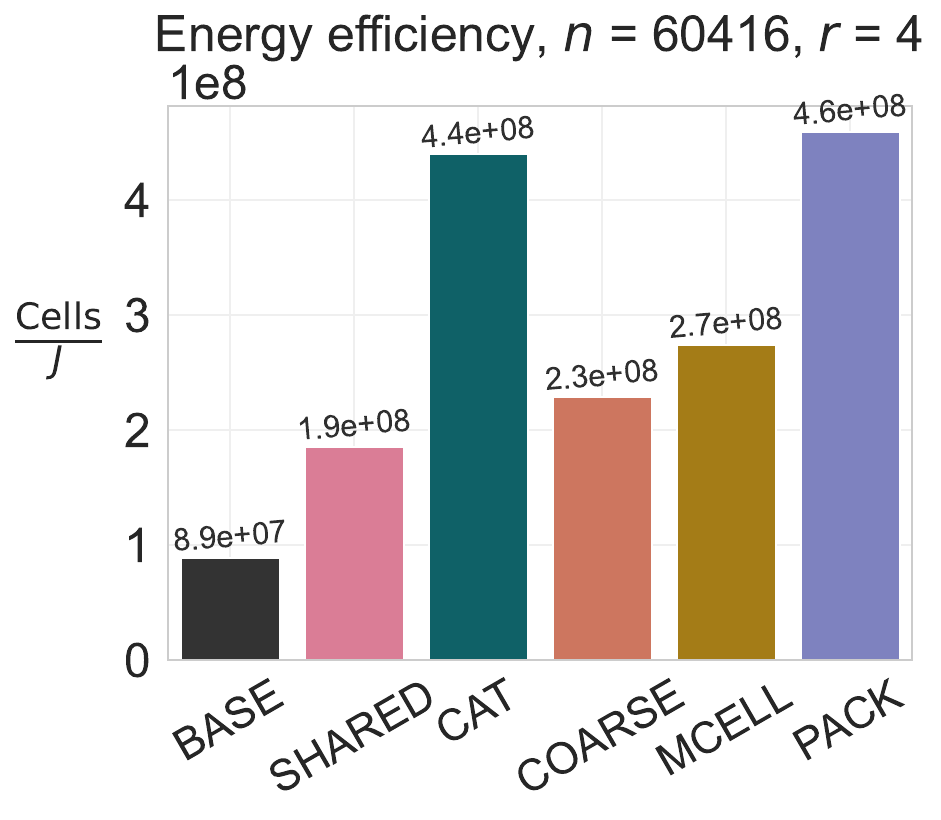}
    \includegraphics[width=0.24\textwidth]{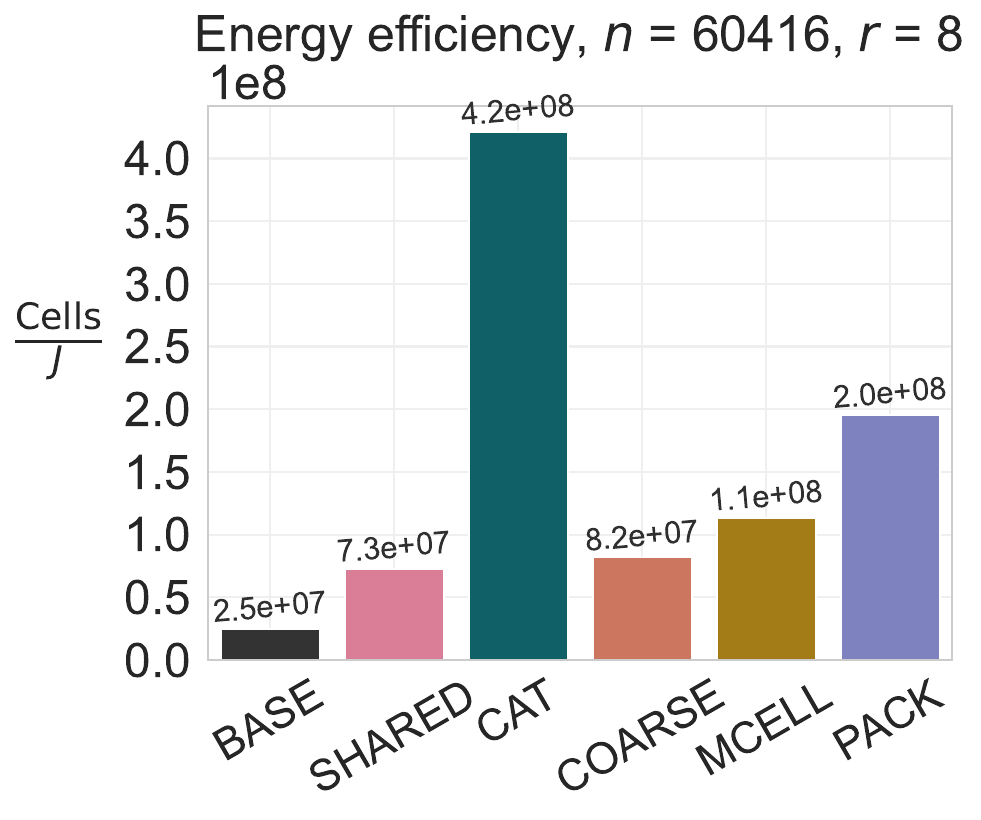}
    \includegraphics[width=0.24\textwidth]{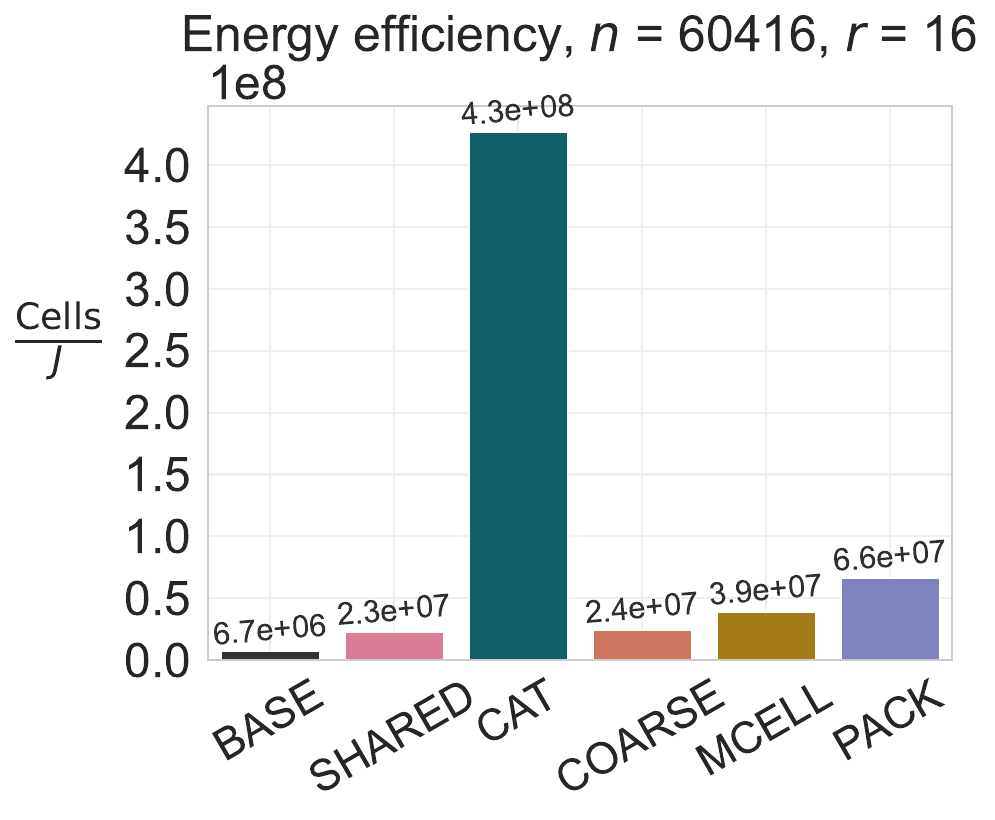}
    \caption{Power time series and energy efficiency for all approaches at different values of neighborhood $r$.}
    \label{fig:energy}
\end{figure*}

Figure \ref{fig:energy} presents the power consumption and energy efficiency of all approaches, for radiuses $r=1,4,8,16$. For the power consumption plots, CAT's energetic behavior is a short high-power curve that peaks near $700\text{W}$ which is the TDP of the H100 GPU. At $r=1$ CAT's energy efficiency is in the same range of the other approaches, except for PACK that is much more energy efficient by a great margin. This can be explained by the bit-level logic of PACK which is highly efficient. However as $r$ increases, CAT's power consumption curve remains roughly the same, while the other approaches sustain their consumption for longer thus using more energy. This difference is most noticeable at $r=16$ where CAT can be up to $6.45\times$ more energy efficient than the second best, and an order of magnitude more efficient than the rest. For the other approaches, their power curves tend to transform from pulse-like shapes to plateau-like ones.    

\begin{figure}[ht!]
    \centering
    \includegraphics[width=0.24\textwidth]{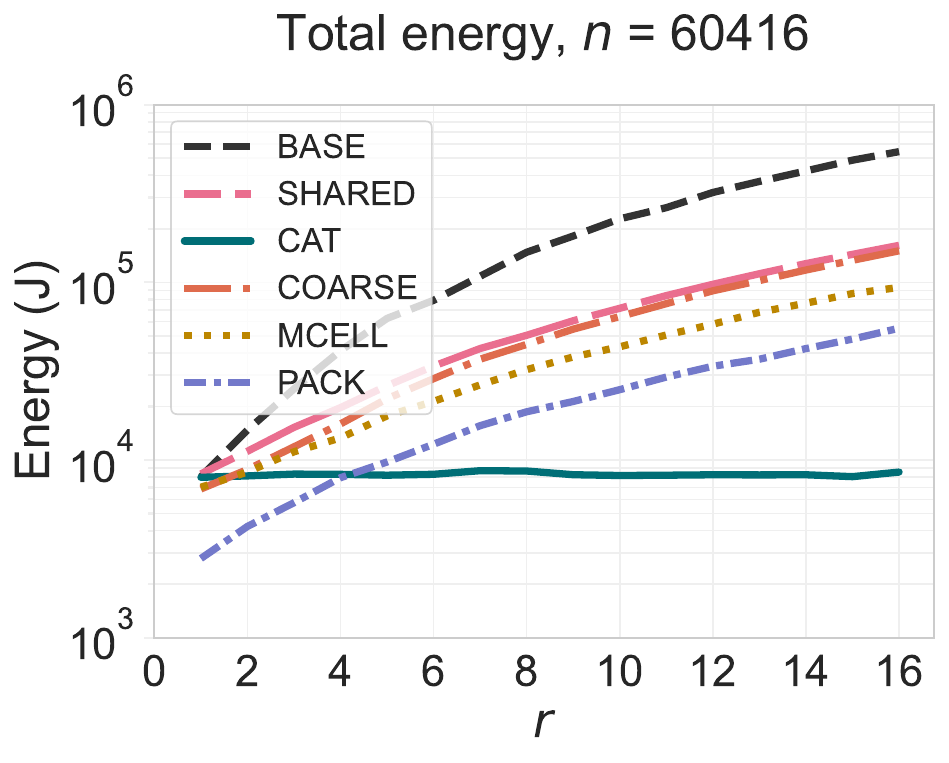}
    \includegraphics[width=0.24\textwidth]{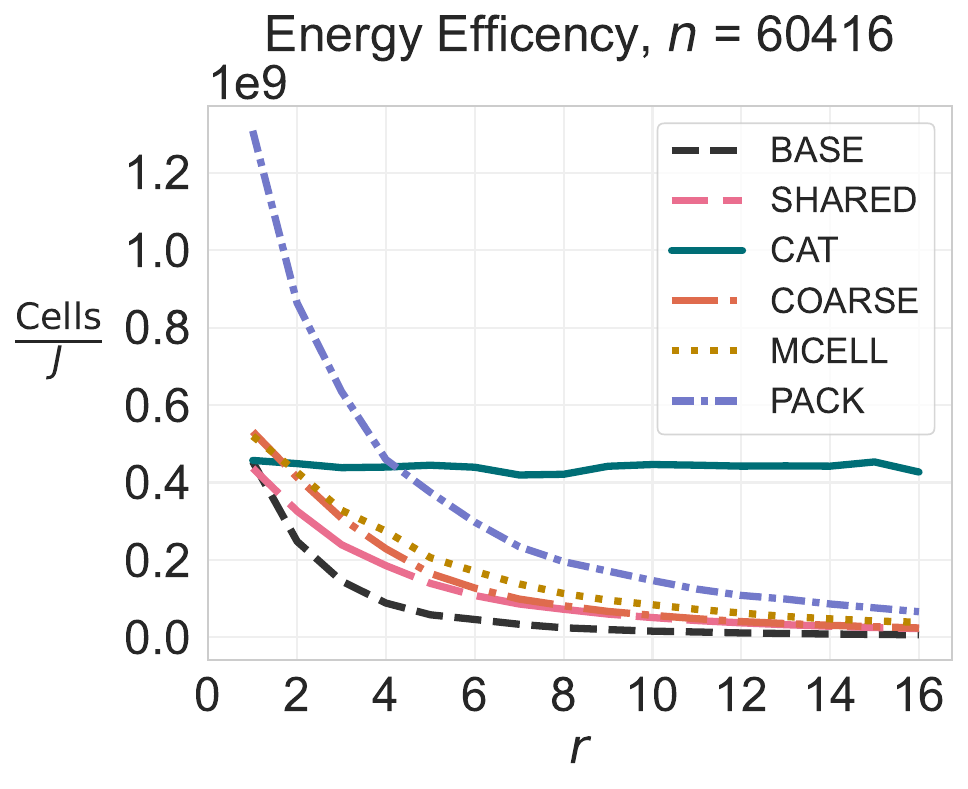}
    \caption{Impact of the neighborhood radius $r$ in the total energy and energy efficiency for a large CA of $n \times n = 60416 \times 60416$ cells.}
    \label{fig:energy-radius}
\end{figure}

Figure \ref{fig:energy-radius} shows more in detail how increasing $r$ affects the total energy used and energy efficiency of each approach. CAT's total energy is remains almost unchanged across the entire range of $r$, while the other approaches use more energy with higher $r$. In terms of energy efficiency (Cells per Joule), $r=[4..5]$ is the crossover zone where CAT's surpasses PACK and becomes the most energy efficient approach. Below this value, PACK is the most energy efficient one. As for the other approaches in general they are more energy efficient than BASE, although their differences are smaller at high $r$. 

\subsection{Scaling with GPU Generations}
The first generation of tensor cores was introduced back in 2017 with NVIDIA's V100 GPU, which had the Volta architecture. Since then, different tensor core generations were released, with improved performance and features. The second generation of tensor cores was introduced in 2018 with the Turing architecture, focused on videogames (e.g., the TITAN RTX GPU) with no data-center variant. The third generation released in 2020 with the Ampere architecture (e.g., the A100 GPU), and the fourth generation in 2022 with the Hopper architecture (e.g., the H100 GPU). Many other aspects also got improved on each generational jump, such as the number of traditional cores (\texttt{FP32/INT32} units), memory capacity, memory bandwidth, among others. These improvements make both classical and tensor core based GPU implementations to automatically scale their performance by just using newer hardware. The scaling factors provided by the last GPU architectures can give key insights on what performance could future GPU architectures bring to CAT in comparison to the other approaches.

\begin{figure*}[ht!]
    \centering
    \includegraphics[width=0.24\textwidth]{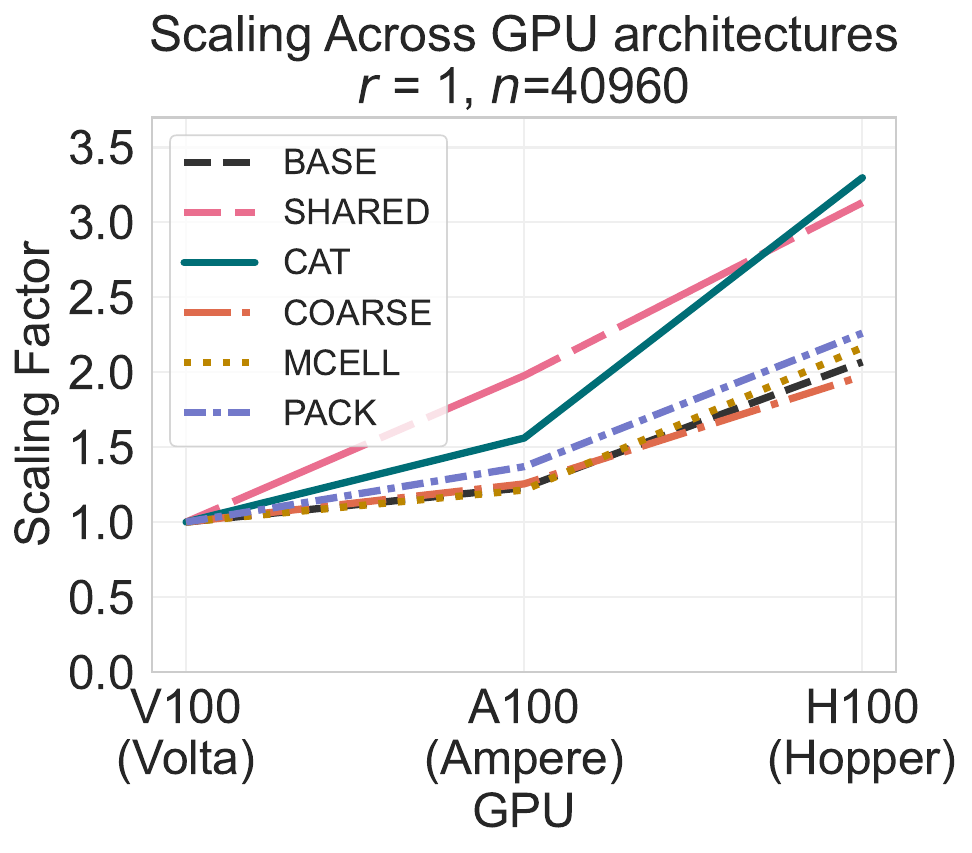}
    \includegraphics[width=0.24\textwidth]{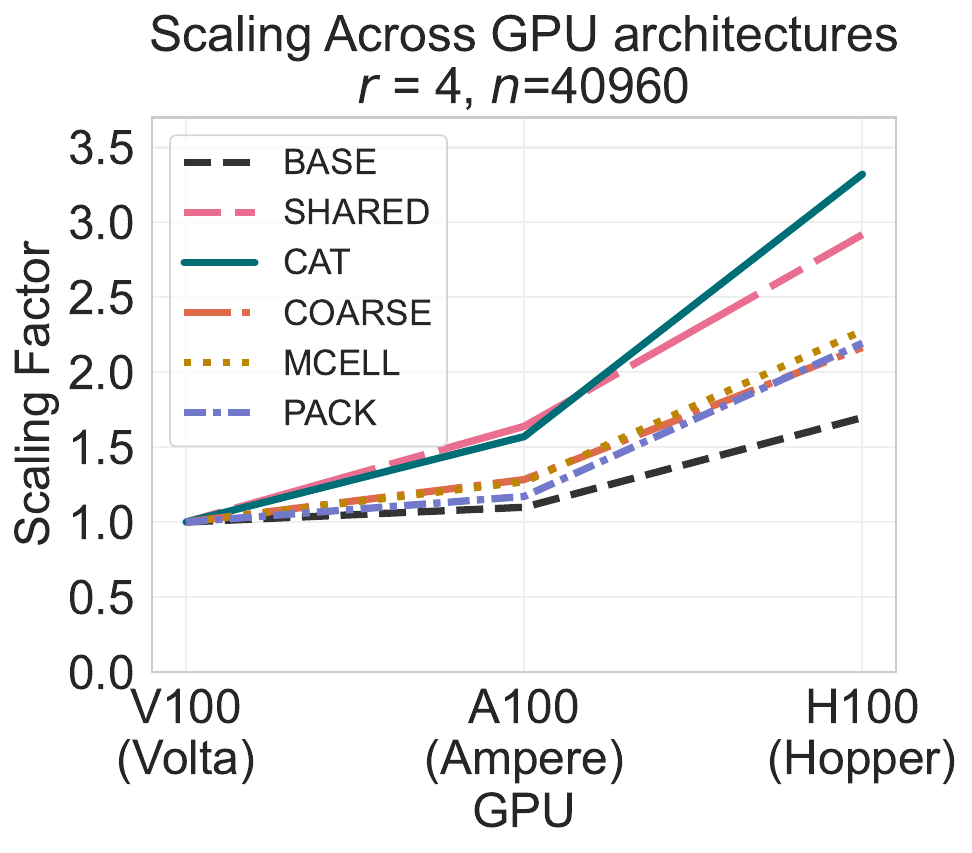}
    \includegraphics[width=0.24\textwidth]{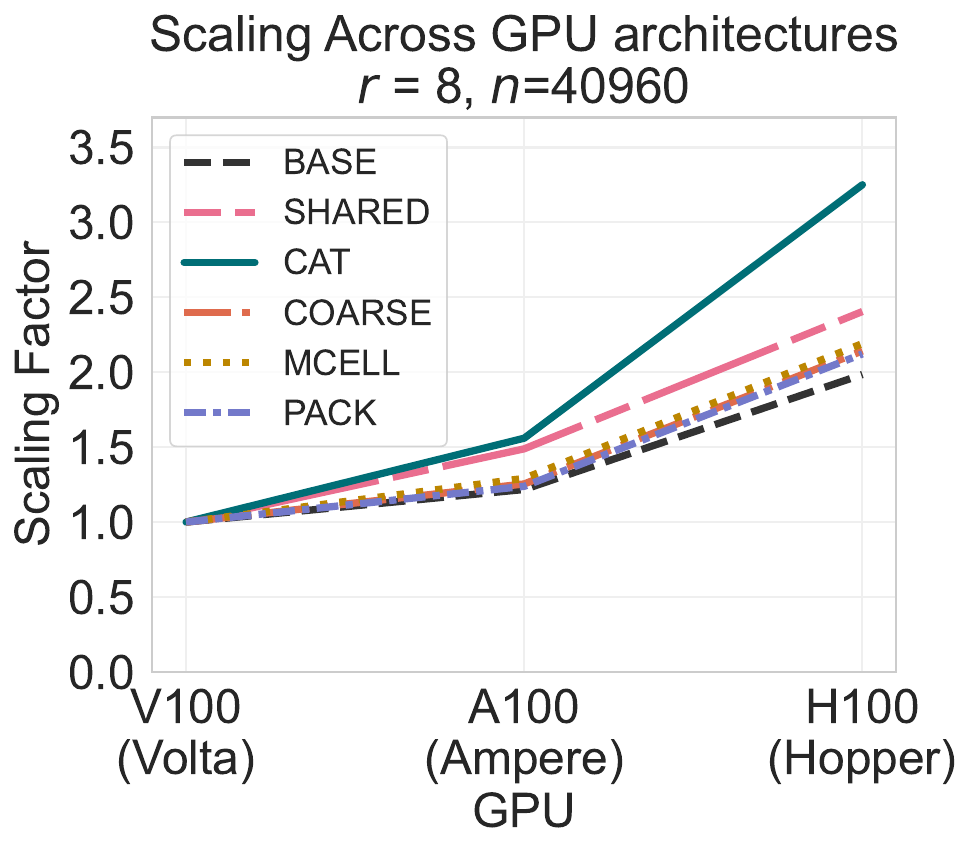}
    \includegraphics[width=0.24\textwidth]{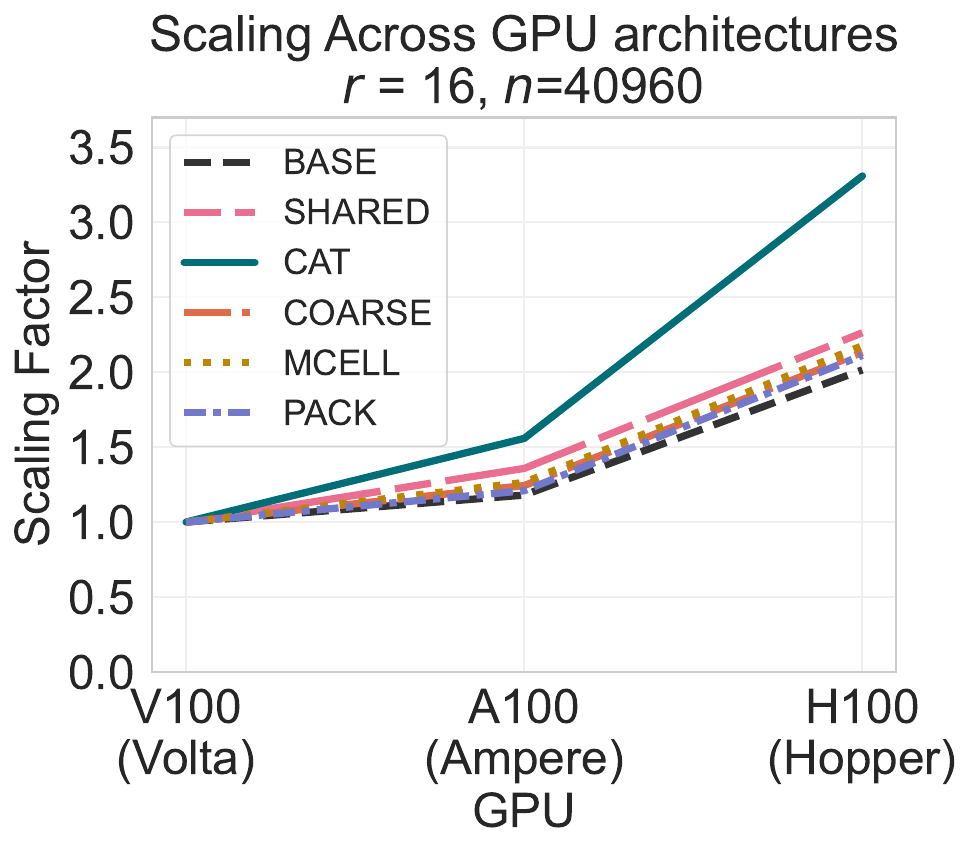}
    \caption{Scaling factors of CAT and all other approaches with respect to their performance on the Volta V100.}
    \label{fig:scaling-factors}
\end{figure*}

Figure \ref{fig:scaling-factors} presents the scaling factors of CAT and all other approaches, relative to their performance on a Volta V100 GPU. Overall, CAT shows the highest scaling factors across GPU architectures, specially when switching from the A100 to the H100 where it achieves a scaling of near $3.3\times$ in all cases. A particular behavior is noted at $r=1$ where SHARED exhibits the highest scaling when jumping from the V100 to the A100. But in general, the transition from Ampere to Hopper is the one that provides the highest jump, and tensor core performance has been scaling at a higher rate than regular GPU compute. With the field of AI becoming more relevant each year, it is very likely that this scaling trend could continue, favoring CAT with a scaling rate that is higher than traditional GPU approaches.

\section{Discussion and Conclusions}
\label{sec:conclusions}
This work presented CAT (\textbf{C}ellular \textbf{A}utomata on \textbf{T}ensor Cores), a GPU approach that uses tensor cores to accelerate the simulation of cellular automata (CA) at large neighborhood radius. Because of its design based on computing the neighborhood reduction via Matrix Multiply Accumulate (MMA) operations, CAT's performance is unaffected by the increase in neighborhood radius in the range $1 \le r \le 16$, while other state of the art approaches do increase their work (hence time) as $r$ increases. CAT can be employed on any CA model where the cell transition function acts over a weighted summation of its neighbors, such as the life-like CA models. 

Experiments using the Larger Than Life (LTL) family of CA as case studies showed that CAT is the fastest approach of all in the range $3 \le r \le 16$; reaching up to $101\times$ of speedup over a traditional baseline approach and up to $\sim14\times$ faster than the fastest state of the art GPU method (PACK). For the low range $1 \le r \le 2$, CAT is still not the fastest approach but still competitive only being behind the GPU packet coding approach (PACK) which is efficient at low radius. In terms of energy, CAT is the most energy efficient approach in the range $5 \le r \le 16$; up to $6.45\times$ more efficient than second most efficient approach (PACK), and up to an order of magnitude more efficient than the rest.  

In terms of performance scaling across GPU architectures, the last two tensor core generations (Ampere and Hopper) have provided significant scaling factors to CAT which are much higher than the ones observed for the other approaches. If we assume that at least part of this trend continues in the future, supported by the fact that AI is driving the evolution of GPUs towards more tensor core performance, then CAT becomes a promising approach for upcoming GPU architectures.

As for future work, CAT can be further improved and extended in several aspects. For instance, caching the tiles of $\Lambda$ into \textit{shared memory} should make CAT run even faster; a preliminary version was attempted but reported slower performance than the actual version of CAT. Further research and experimentation is required to reach a proper caching scheme of the $\Lambda$ tile. Another technical improvement for CAT is to support more neighborhood types; currently CAT's design supports \textit{Moore} and \textit{Simplified Von Neumann} neighborhoods. A third neighborhood that is often required is the original \textit{Von Neumann} neighborhood, which has a diamond shape. Supporting this neighborhood would require re-defining the non-zero entries of the band matrices to capture the diamond-shaped neighborhood, as well as verifying if the transition from step 1 to step 2 of MMAs should accumulate or multiply the previous result. A third technical improvement is to take advantage of the low precision types currently supported by tensor cores, such as \texttt{INT8} which is faster than the \texttt{FP16} types currently used, or the experimental types \texttt{BIT},\texttt{INT4}, which are even faster. Some of these types are only supported by rectangular shaped fragments, but the square shaped requirement can still be accomplished by stacking several of these fragments until a major square region is built again. Currently, what penalizes this last idea is the need of converting all \texttt{INT32} results back again into low precision types in between the two major steps of CAT. Finding a faster way of re-using the resulting \texttt{INT32} fragment back as a product operand for another MMA could indeed accelerate CAT significantly. 

CAT's neighborhood radius range can be extended beyond $1 \le r \le 16$; it only requires to expand the main idea presented in Figure \ref{fig:cat-overview} to consider the next closest fragments at each reduction direction, \textit{i.e.}, $F_{i\pm 2,j \pm 2}$. This expansion implies having six band fragments $\pi_1, \pi_2, \dots, \pi_6$ instead of three, as well as to widen the global halo in one more fragment. All of these changes would make CAT support neighborhood radiuses in the range $1 \le r \le 32$.   
%but this is not necessarily a problem 
%since LTL rules only need a lot of precision when the amount of neighbors is close to the rule limit. Thus, this problem 
%and could be avoided by applying a transformation to the rule such that it can also be applied to each fragment before combining them. 
In general, successive applications of this expansion would make CAT increase its supported radius range by $+16$. Each new expansion would put CAT's running time at a higher plateau of running time, but would remain constant for that new supported range. The challenge however is that \texttt{FP16} may not be precise enough to store the maximum amount of neighbors for such large radiuses. Possible solutions could be to normalize the range of values, which would push the limitation to appear at a larger $r$, or a more fundamental solution is that future GPUs could support full \texttt{INT32} MMAs.

Lastly, exploring the 3D extension of CAT is an natural extension to continue researching on. The challenge would be to find ways to represent 3D reductions with tensor cores that operate on 2D MMAs. Reusing the main idea of CAT, now by layers is feasible, but also other new approaches could be explored and tested, specially given that generic arithmetic reductions are known to run efficient on tensor cores \cite{navarro2020gpu}, opening the possibility to reduce entire volumes of cells with a few MMAs.

To conclude, this work has shown that tensor cores can indeed accelerate non-AI applications such as the simulation of cellular automata. In particular, CAT can be of great interest for research teams that need to study complex phenomena that emerges from CA with large neighborhood radius.

\section*{Acknowledgment}
This work was supported by the ANID FONDECYT grants \#1221357, \#1241596, the \text{Temporal} research lab, and the Patag\'on Supercomputer of Austral University of Chile (FONDEQUIP EQM180042). The authors thank Roberto Melita, who brought rich ideas and discussions in the early stages of this research.

\bibliographystyle{IEEEtran}
\bibliography{main}

\end{document}